\documentclass[11pt]{article}
\usepackage[margin=1in]{geometry}
\usepackage{setspace}
\usepackage{dsfont, amssymb,amsmath,amscd,latexsym, amsthm, amsxtra,amsfonts, extarrows}
\usepackage{amssymb}
\usepackage{amsmath}
\numberwithin{equation}{section}
\numberwithin{figure}{section}
\usepackage{indentfirst}
\usepackage{graphicx}
\usepackage[colorlinks=true, allcolors=blue]{hyperref}
\usepackage[utf8]{inputenc} 
\usepackage[T1]{fontenc}    
\usepackage{url}            
\usepackage{booktabs}       
\usepackage{amsfonts}       
\usepackage{nicefrac}       
\usepackage{microtype}      
\usepackage{fancyhdr}       
\usepackage{natbib}
\graphicspath{{media/}}

\newtheorem{Remark}{Remark}

\newtheorem{Proposition}{Proposition}
\newtheorem{Assumption}{Assumption}
\usepackage{appendix}
\usepackage{subcaption}

\usepackage{afterpage}
\usepackage{rotating}

\begin{document}

\title{\textbf{Robust Investment-Driven Insurance Pricing and Liquidity Management}} 

\author{Bingzheng Chen\thanks{\ School of Economics and Management, Tsinghua University; Email: chenbzh@sem.tsinghua.edu.cn} \and Jan Dhaene\thanks{\ Faculty of Economics and Business, Katholieke Universiteit Leuven; Email: jan.dhaene@kuleuven.be} \and Chun Liu\thanks{\ School of Economics and Management, Tsinghua University; Email: liuch@sem.tsinghua.edu.cn} \and Shunzhi Pang\thanks{\ School of Economics and Management, Tsinghua University; Faculty of Economics and Business, Katholieke Universiteit Leuven; Email: psz22@mails.tsinghua.edu.cn}}

\date{\today}
\maketitle

\begin{abstract}
\noindent 
This paper develops a dynamic equilibrium model of the insurance market that jointly characterizes insurers' underwriting, investment, recapitalization, and dividend policies under model uncertainty and financial frictions. Competitive insurers maximize shareholder value under a subjective worst-case probability measure, giving rise to liquidity-driven underwriting cycles and flight-to-quality behavior. Model uncertainty acts as an informational friction on insurers' risk-taking behavior and helps regularize a finite-barrier verification system in settings with external financial investment opportunities. We further show that robustness concerns do not eliminate the investment-hedging channel in insurance pricing: when underwriting surplus and financial returns are sufficiently negatively correlated, the hedging value of financial investment can be passed through to policyholders, leading to lower insurance prices and, in high-capacity states, negative equilibrium loadings. Thus, underwriting losses may arise endogenously even when insurers price rationally under model uncertainty, rather than necessarily reflecting mispricing or irrational underwriting behavior. 
\\
\vspace{0in}\\
\noindent\textbf{Keywords:} Investment-driven insurance pricing; Model uncertainty; Liquidity management; Underwriting cycles; Flight-to-quality; Ergodicity \\ 
\bigskip
\end{abstract}

\newpage

\section{Introduction} 

In recent years, supply-side perspectives have become increasingly important for understanding the functioning of insurance markets. On the one hand, the post-pandemic era, together with rapid advances in artificial intelligence and related technologies, has exposed insurers to new and increasingly complex sources of risk \citep{feng2022editorial, scholes2025artificial}. Owing to limited historical data and incomplete understanding of these emerging risks, many remain difficult to price or manage through transfer, diversification, or hedging. Consequently, insurers are often forced to retain such risks on their balance sheets or even withdraw coverage entirely. 

On the other hand, insurers behave like financial intermediaries: they can raise funds at relatively low cost through underwriting and actively participate in capital markets. For instance, in the U.S. corporate bond market, insurers have long been among the largest institutional investors, playing a crucial role in providing long-term financing to the real economy \citep{koijen2023understanding}. However, empirical evidence suggests that insurers also face substantial financial frictions, particularly in the aftermath of the 2008 global financial crisis, which have imposed shadow costs on insurance pricing and capital allocation \citep{koijen2015cost, koijen2016shadow, ge2022financial}. 

Understanding how supply-side product and financial market frictions affect the equilibrium outcomes of the insurance industry is important, because in a competitive market, these frictions are ultimately passed on to households, thereby influencing overall social welfare. In this paper, we focus on a specific form of informational friction, model uncertainty, also known as ambiguity \citep{knight1921risk, ellsberg1961risk}, which arises from insurers’ limited understanding of the true distribution of risks they face. Early survey evidence among insurers and actuaries has confirmed that risks with ambiguous probabilities tend to be priced much higher \citep{hogarth1992pricing, kunreuther1993insurer}. However, the extent to which such shadow cost should be incorporated into pricing remains a theoretical question, requiring a structural framework to address. Beyond underwriting risks, insurers are also exposed to financial risks, which are widely recognized to exhibit a high degree of ambiguity in their return distributions \citep{epstein2010ambiguity, guidolin2013ambiguity}, further complicating insurers’ pricing and portfolio decisions.

To address these challenges, this paper develops a continuous-time equilibrium model of the insurance market that jointly characterizes insurers’ investment-driven pricing and liquidity management under model uncertainty. On the one hand, we build on the barrier-type equilibrium framework of \citet{henriet2016dynamics}, in which the aggregate liquid reserves held by the industry serve as the state variable governing insurers’ strategies. When reserves rise to a sufficiently high level, insurers distribute dividends to shareholders until capacity falls below the payout boundary; conversely, when reserves decline below a lower threshold, they raise external capital to restore liquidity to the financing boundary. Between these two thresholds lies an internal financing region, within which the equilibrium insurance price, investment amount, and market-to-book ratio are all endogenously determined as functions of aggregate capacity. 

On the other hand, we incorporate insurers’ concerns about model uncertainty following the robust control framework of \citet{hansen2001robust, hansen2008robustness} and \citet{anderson2003quartet}. In this setting, insurers optimize their decisions under the worst-case probability measure subject to an entropy penalty, which captures the trade-off between maximizing firm value and guarding against model misspecification. The resulting worst-case drift distortion can be interpreted as the market price of model uncertainty, representing the implicit shadow cost ultimately borne by policyholders. Compared with the benchmark equilibrium without model uncertainty in \citet{henriet2016dynamics}, introducing robustness concerns leads to a fundamental change in the structure of the equilibrium equations for both the insurance price and the market-to-book ratio. Specifically, the recapitalization boundary is no longer zero but becomes strictly positive, preventing the industry from falling into a liquidity trap. Intuitively, if insurers’ aggregate capacity were to collapse to zero, underwriting and investment activities would cease, and the market could not spontaneously recover. Under robustness, both the equilibrium price and the market-to-book ratio are generally higher, reflecting insurers’ demand for additional compensation for model uncertainty.  

This study is closely related to three papers in the literature, to which we contribute by offering extensions and marginal improvements from distinct perspectives. First, we extend \citet{pang2026robust} by incorporating outside financial investment opportunities, making the model more aligned with real-world practice. Relative to the benchmark without financial investment, the numerically solved equilibrium outcomes show that the minimum, maximum, and overall range of aggregate capacity all expand, indicating that the insurance market can sustain larger cyclical fluctuations. Consequently, the expected duration of underwriting cycles becomes substantially longer. While the market-to-book ratio increases markedly, the equilibrium insurance price does not necessarily rise, reflecting a non-monotonic interaction between pricing and financial exposure. Furthermore, the share of risky investment increases monotonically with insurers’ aggregate liquid reserves, suggesting decreasing effective risk aversion as liquidity buffers accumulate. 

Second, while \citet{luciano2022fluctuations} extend the dynamic equilibrium framework of \citet{henriet2016dynamics} by incorporating financial market exposure and introducing the concept of insurers’ risk appetite, we build upon theirs by explicitly introducing concerns for model uncertainty, thereby capturing a form of supply-side informational friction. We further generalize their setup by allowing for correlation between insurance and financial market risks, rather than assuming independence. More importantly, we would like to gently point out that the equilibrium defined in \citet{luciano2022fluctuations} does not, in fact, theoretically exist, at least not in a barrier-type form. The reason is that, once outside financial investment is introduced, the aggregate investment rule becomes ill-defined at the payout boundary. At such a boundary, the implied market risk aversion vanishes, while the risky financial asset still carries a positive Sharpe ratio. Insurers therefore have an incentive to allocate additional liquid reserves to financial investment rather than distribute them as dividends, making it impossible to sustain a finite payout boundary under their formulation. 

Model uncertainty changes this logic by adding an informational friction to financial risk-taking. Large investment positions expose insurers not only to financial volatility, but also to concerns about model misspecification. This restrains the demand for risky investment and helps regularize the finite-barrier verification system. We do not claim that such a robust equilibrium exists for all parameter values. Instead, we characterize a candidate stationary Markovian equilibrium through a verification system and show numerically that admissible solutions arise under economically reasonable parameterizations.

Third, we examine how the correlation between insurance and financial market risks affects insurance pricing, and show that the negative-loading phenomenon highlighted by \citet{chen2025dynamic} can also arise under a robust pricing regime. This issue is important for understanding the increasingly common occurrence of underwriting losses in insurance markets.\footnote{According to a 2022 research report by the China Insurance Security Fund, over half of P\&C insurers in China have experienced underwriting losses for five consecutive years.} \citet{chen2025dynamic}, which also builds on the equilibrium pricing framework of \citet{henriet2016dynamics}, studies the joint underwriting-investment problem of a representative CARA insurer and shows that the endogenous loading factor may become negative when underwriting gains and financial returns are negatively correlated, because investment income can hedge underwriting losses. By contrast, when the insurance and financial markets are independent or positively correlated, the equilibrium loading remains strictly positive. We find a similar pattern in our robust equilibrium: when underwriting surplus and financial returns are negatively correlated, the equilibrium loading can turn negative in high-capacity states. Thus, although model uncertainty introduces an additional informational friction, it does not eliminate the underwriting-investment hedging incentive. This result suggests that observed underwriting losses in reality need not be interpreted simply as pricing mistakes; they may also arise from an equilibrium investment-driven pricing mechanism.  

Finally, we use the equilibrium characterization to study insurers' long-run behavior. Because aggregate liquid reserves are reflected between the recapitalization and payout boundaries, the model generates endogenous underwriting cycles: when capacity is high, insurance prices decline and the market enters a soft phase; when adverse shocks reduce capacity, prices rise and the market moves into a hard phase. Financial investment also varies over the cycle. In low-capacity states, insurers reduce their exposure to risky assets, consistent with the flight-to-quality behavior documented in insurance markets \citep{luciano2022fluctuations}. We further characterize the reflected capacity process and its invariant distribution under the worst-case measure, which allows us to discuss the ergodic properties of the insurance market. This analysis links equilibrium pricing, liquidity management, investment behavior, and long-run market fluctuations within a unified framework.  

To summarize, our paper contributes to the growing literature on the financial economics of insurance, which emphasizes extending the supply-side theory of insurance markets by incorporating financial frictions, regulatory constraints, and imperfect competition \citep{koijen2023financial}.\footnote{Related studies include \citet{koijen2015cost, koijen2016shadow, koijen2022fragility, koijen2023understanding, koijen2024aggregate}, \citet{ge2021role, ge2022financial}, \citet{egan2022conflicting}, \citet{ellul2022insurers}, and \citet{kubitza2021investor}.} Following the idea that equity is costly for financial intermediaries \citep{he2013intermediary, brunnermeier2014macroeconomic}, we treat equity financing costs as the key form of financial friction faced by insurers. This mechanism makes liquidity management the central determinant of insurers’ dynamic equilibrium behavior, giving rise to the underwriting cycles and flight-to-quality phenomena. In addition, we highlight the informational friction arising from uncertainty in the loss distribution, which plays an essential role in insurers’ exposure to emerging risks such as climate and cyber risks \citep{koijen2022new}. We argue that robust pricing frameworks, which explicitly account for model uncertainty, should become an integral part of modern insurance practice. 

We also complement the classical literature in actuarial science and insurer risk management. From a pricing perspective, we build on the equilibrium pricing framework, which emphasizes that insurance prices depend not only on underlying risk characteristics but also on market conditions and capital constraints \citep{buhlmann1980economic, buhlmann1984general, winter1994dynamics, henriet2016dynamics, feng2023insurance}. From a risk management perspective, while the implications of ambiguity have been extensively examined in the context of insurers’ investment and reinsurance strategies,\footnote{Related studies include \citet{zeng2016robust}, \citet{li2019optimal}, \citet{sun2019robust}, among others.} its direct impact on insurance pricing has received comparatively less attention despite its first-order importance. Recent studies, such as \citet{zhao2011ambiguity}, \citet{anwar2012competitive}, \citet{dietz2019ambiguity}, and \citet{dietz2021pricing}, find that ambiguity aversion tends to increase insurance prices. In contrast, the advantage of our structural framework lies in its ability to quantify the magnitude of these price adjustments, thereby providing theoretical guidance for practical pricing and policy design. 

The remainder of the paper is organized as follows. Section~\ref{Section Model} introduces the theoretical framework. Section~\ref{Section Equilibrium} solves the robust optimization problem and presents the main equilibrium results. Section~\ref{Section Numerical} provides numerical analyses of the equilibrium outcomes. Section~\ref{Section Long-Run Behavior} examines insurers’ long-run dynamics. Finally, Section~\ref{Section Conclusion} concludes the paper. 

\section{Theoretical Framework} \label{Section Model}

This section develops the theoretical model by extending the framework of \citet{luciano2022fluctuations} to account for insurers’ concerns about model uncertainty, following the robust control approach of \citet{hansen2001robust} and \citet{anderson2003quartet}. Consider a filtered probability space $(\Omega,\mathcal{F},\{\mathcal{F}_t\}_{t\geq0},\mathbb{P})$ satisfying the usual regularity conditions. All stochastic processes governing the insurance and financial markets below are assumed to be well-defined on this space. 

\subsection{Insurance Market} 

We consider a competitive insurance market composed of a continuum of firms, each offering coverage for perfectly correlated aggregate risks. Idiosyncratic components at the individual level are ignored, as they can be fully diversified across a sufficiently large pool and therefore do not affect market-level equilibrium dynamics. The focus of the model is thus on systematic risk that cannot be diversified within the insurance sector.\footnote{A similar setting, where insurers face primarily aggregate or systemic risk, has been widely adopted in the literature \citep{henriet2016dynamics, feng2023insurance, chen2025dynamic}. \citet{luciano2022fluctuations} explicitly incorporates firm-specific idiosyncratic risk. However, due to diversification across a large number of policies, such idiosyncratic components do not materially affect the equilibrium outcomes. Therefore, although we omit them here for simplicity, our framework remains directly comparable to that of \citet{luciano2022fluctuations}.}  The cumulative loss process for a representative insurer, denoted by $L \triangleq \Big\{ L_t: t \geq 0 \Big\}$, follows the stochastic evolution: 
\begin{equation}
    \mathrm{d}L_t = l \mathrm{d}t - \eta \mathrm{d}W^{I}_t, \label{Benchmark Loss Process}
\end{equation}
where $l > 0$ represents the expected instantaneous loss intensity, $\eta > 0$ measures exposure to systematic loss shocks, and $W^I \triangleq \Big\{ W^I_t: t \geq 0 \Big\}$ is a standard Brownian motion.\footnote{In practice, aggregate loss risk may arise from multiple sources (e.g., macroeconomic shocks, natural catastrophes, or climate-related events) and can be modeled by a multi-dimensional Brownian motion. For analytical tractability, we aggregate these sources into a single risk factor. } 

Insurance contracts are assumed to be short-term. Such setting, commonly adopted in the literature on dynamic insurance markets \citep[e.g.,][]{henriet2016dynamics, luciano2022fluctuations}, is particularly appropriate for the property and casualty (P\&C) sector, where policies typically have short maturities and insurers do not carry long-term liabilities. The insurance premium per unit time is specified as: 
\begin{equation}
    \pi_t \mathrm{d}t = (l + \eta p_t) \mathrm{d}t, \label{Premium}
\end{equation}
which consists of two components: an actuarially fair premium $l$, and a loading premium, depending on the risk exposure $\eta$ and a loading factor $p_t$. The loading factor is endogenously determined in equilibrium through market clearing between aggregate supply and demand. 

While insurance demand can be modeled in various ways, we follow the approach of \citet{luciano2022fluctuations} and endogenize the demand function within a simple general-equilibrium framework. Suppose a representative insuree faces the unit loss process in \eqref{Benchmark Loss Process} and purchases insurance at premium $\pi_t$. Let $d_t$ denote the quantity of coverage demanded for the normalized unit risk. Thus, $d_t=1$ corresponds to full coverage of the benchmark unit exposure, while $1-d_t$ is the retained exposure. Under mean-variance preferences over an infinitesimal interval $\mathrm{d}t$, the optimal coverage is given by $d_t^\ast=1-\frac{1}{\alpha\eta}p_t$, where $\alpha>0$ denotes the degree of risk aversion.\footnote{The full derivation appears in Appendix A of \citet{luciano2022fluctuations}. For analytical simplicity, we do not impose the constraint $d_t\in[0,1]$ at this stage.} Hence, aggregate demand can be written as
\begin{equation}
    D(p)=1-\frac{1}{\alpha\eta}p. \label{demand function}
\end{equation}
When $p=0$, the representative insuree demands full coverage of the normalized unit risk.

Several remarks are useful for interpreting this demand specification. First, $D(p)$ should be understood as demand for normalized risk units, rather than as a physical constraint on any individual insurer's underwriting share. Second, the linear form is adopted for tractability and to maintain comparability with \citet{luciano2022fluctuations}. Since the focus of the paper is on insurers' supply-side pricing, investment, and liquidity-management decisions, the specific functional form of demand is not central to the mechanism; similar to \citet{henriet2016dynamics}, a general downward-sloping demand curve would change the algebra but not the role of aggregate capacity and financial investment in determining equilibrium prices. Finally, because the constraint $d_t\in[0,1]$ is not imposed in the interior analysis, a negative loading may imply $D(p)>1$. This case should not be interpreted as literal over-insurance of a fixed individual loss, but rather as demand for more than one unit of the normalized risk exposure when insurance is sufficiently inexpensive. Imposing a bounded coverage constraint would introduce corner regions in demand, whereas the interior pricing mechanism studied below would remain unchanged.

Since insurance contracts are assumed to be of infinitesimal duration, insurers hold no long-term liabilities and thus are not required to maintain reserves against future claims. On each insurer’s balance sheet, the liquid reserves on the asset side, denoted by $m \triangleq \Big\{m_t: t \geq 0 \Big\}$, must equal the equity position $e \triangleq \Big\{e_t: t \geq 0 \Big\}$. Accordingly, $m_t$ simultaneously represents the book value of equity and the insurer’s net wealth. 

Suppose an insurer’s underwriting scale process is denoted by $x \triangleq \Big\{x_t \geq 0: t \geq 0 \Big\}$. Over an infinitesimal interval $(t, t + \mathrm{d}t)$, the insurer receives premium income of $x_t \pi_t \mathrm{d}t$ and pays out claims of $x_t \mathrm{d}L_t$. The difference $x_t \big( \pi_t \mathrm{d}t - \mathrm{d}L_t \big)$, represents the underwriting profit, which can either be retained as liquid reserves or distributed as dividends. Let $\delta \triangleq \Big\{\delta_t: t \geq 0 \Big\}$ denote the cumulative dividend process, where the insurer optimally chooses the payout $\mathrm{d}\delta_t \geq 0$ to distribute to shareholders. In the event of financial distress, the insurer may also raise external capital to restore liquidity. Let $i \triangleq \Big\{i_t: t \geq 0 \Big\}$ denote the cumulative recapitalization process, with $\mathrm{d}i_t \geq 0$. The insurer’s net wealth process therefore evolves as: 
\begin{equation}
    \mathrm{d} m_t = x_t \big( \pi_t \mathrm{d}t - \mathrm{d}L_t \big) + \mathrm{d} i_t - \mathrm{d}\delta_t = x_t \eta p_t \mathrm{d}t + x_t \eta \mathrm{d}W^I_t + \mathrm{d}i_t - \mathrm{d}\delta_t. \label{Benchmark Individual Wealth Process}
\end{equation}

External financing is generally costly for firms \citep{jensen1976theory, leland1977informational, myers1984corporate}. The same holds for financial intermediaries such as insurers, where regulatory capital requirements and solvency constraints further amplify financing costs \citep{cummins1994capital, cummins2005estimating}. Let $\gamma > 0$ denote the per-unit cost of raising external capital, which is incorporated into the insurer’s financial valuation. Finally, we assume that all insurers begin with strictly positive liquid reserves, i.e., $m^j_0 > 0$ for all $j \in \mathcal{J}$.\footnote{This assumption guarantees that, in equilibrium, all insurers’ liquid reserves remain of the same sign at any point in time, as will be shown later.}

\subsection{Financial Investment Opportunity}

We now extend the setting to allow insurers to invest in financial markets alongside their underwriting activities. The capital market follows a standard Black–Scholes framework with a risk-free asset (e.g., a bond) and a risky asset (e.g., a stock or market portfolio). The asset price dynamics are described by the following stochastic differential equations (SDEs): 
\begin{equation}
    \left\{
    \begin{aligned}
        & \frac{\mathrm{d}B_t}{B_t} = r \mathrm{d}t, \\
        & \frac{\mathrm{d}S_t}{S_t} = \mu \mathrm{d}t + \sigma \mathrm{d}W^S_t, 
    \end{aligned}
    \right. \notag
\end{equation}
where $r$ is the risk-free rate, $\mu \geq r$ represents the expected return on the risky asset, and $\sigma > 0$ denotes the volatility of the return. Let $q \triangleq \frac{\mu - r}{\sigma} > 0$ denote the Sharpe ratio. Besides, $W^S \triangleq \Big\{ W^S_t: t \geq 0 \Big\}$ is a one-dimensional standard Brownian motion, which may be correlated with the insurance loss risk $W^I$. The correlation coefficient is given by $\rho \in (-1, 1)$, such that $\mathrm{d}W^I_t \mathrm{d}W^S_t = \rho \mathrm{d}t$. In the main analysis of this paper, we exclude the boundary cases $\rho = \pm 1$ to ensure mathematical tractability, as perfect correlation would lead to degeneracies in the stochastic system. 

In this setup, the financial market is assumed to be frictionless, allowing insurers to adjust their portfolios continuously. Consider an insurer with liquid reserves process $m$, which adopts an investment strategy $y \triangleq \Big\{ y_t: t \geq 0 \Big\}$, putting $y_t$ amount to the risky asset at time $t$, and the remaining $m_t - y_t$ to the risk-free asset. The insurer’s net wealth process then evolves as: 
\begin{equation}
    \mathrm{d}m_t = (x_t \eta p_t + y_t \sigma q + m_t r ) \mathrm{d}t + x_t \eta \mathrm{d}W^I_t + y_t \sigma \mathrm{d} W^S_t + \mathrm{d}i_t - \mathrm{d}\delta_t. \label{Benchmark Individual Wealth Process with Investment}
\end{equation} 
This formulation allows for a unified analysis of the insurer’s joint underwriting and investment decisions within a dynamic equilibrium framework. 

\subsection{Optimization Objective}

Let $\mathcal{J} \triangleq [0,1]$ denote the set of insurers, each indexed by $j \in \mathcal{J}$. Given the individual liquid reserve process $m^j$, the aggregate liquid reserves (or capacity) of the entire insurance sector are defined as $M \triangleq \Big\{ M_t = \int_{\mathcal{J}} m^j_t \mathrm{d}j: t \geq 0 \Big\}$, which evolve according to: 
\begin{align}
    \mathrm{d}M_t & = (X_t \eta p_t + Y_t \sigma q + M_t r) \mathrm{d}t + X_t \eta \mathrm{d} W^I_t + Y_t \sigma \mathrm{d} W^S_t + \mathrm{d}I_t - \mathrm{d}\Delta_t \notag \\
    & = \Big(D(p_t) \eta p_t + Y_t \sigma q + M_t r\Big) \mathrm{d}t + D(p_t) \eta \mathrm{d} W^I_t + Y_t \sigma \mathrm{d} W^S_t + \mathrm{d}I_t - \mathrm{d}\Delta_t,  \label{Benchmark Aggregate Wealth Process}
\end{align}
where $X_t = \int_{\mathcal{J}} x^j_t \mathrm{d}j$, $Y_t = \int_{\mathcal{J}} y^j_t \mathrm{d}j$, $I_t = \int_{\mathcal{J}} i^j_t \mathrm{d}j$ and $\Delta_t = \int_{\mathcal{J}} \delta^j_t \mathrm{d}j$, represent the aggregate underwriting scale, investment amount, cumulative recapitalization and dividend payouts, respectively. In equilibrium, aggregate supply must match market demand, leading to the market-clearing condition $X_t = D(p_t)$. We focus on a Markovian stationary equilibrium, in which the insurance price depends deterministically on aggregate capacity, that is, $p_t = p(M_t)$.

Without concerns about model uncertainty, each insurer takes the aggregate dynamics in \eqref{Benchmark Aggregate Wealth Process} and the pricing function $p(M)$ as given, and optimally chooses its underwriting scale $x \ge 0$, investment amount $y$, recapitalization policy $\mathrm{d}i \ge 0$, and dividend policy $\mathrm{d}\delta \ge 0$ to maximize shareholder value: 
\begin{equation}
    v(m, M) = \max_{x \geq 0, y, \mathrm{d}\delta \geq 0, \mathrm{d}i \geq 0} \mathbb{E} \left\{ \int_0^{\infty} e^{-\lambda t} \Big[ \mathrm{d}\delta_t - (1 + \gamma) \mathrm{d} i_t \Big] \right\}, \label{Benchmark Optimization Objective}
\end{equation}
where $\lambda \geq r > 0$ is the discount rate applied to future cash flows. 

We now introduce model uncertainty about the probability law governing the insurance and financial risk factors. Insurers evaluate their decisions against alternative probability measures that are locally equivalent to the reference measure $\mathbb{P}$. Let $W_t=(W_t^I,W_t^S)^\top$ denote the Brownian risk vector, whose instantaneous covariance matrix is
\begin{equation}
C=
\begin{pmatrix}
1 & \rho \\
\rho & 1
\end{pmatrix},
\qquad \rho\in(-1,1). \notag
\end{equation}
We parameterize model misspecification by a progressively measurable drift distortion $h=(h^I,h^S)^\top$. Since the two Brownian components are correlated, the stochastic exponential associated with the distorted measure is written in terms of the covariance-adjusted integrand $C^{-1}h$. Let $\mathcal{H}$ denote the admissible set of distortion processes. For each $h\in\mathcal{H}$, define $\xi^h\triangleq\Big\{\xi_t^h:t\geq0\Big\}$ by
\begin{equation}
\xi_t^h = \exp\left\{ \int_0^t (C^{-1}h_s)^\top \mathrm{d}W_s - \frac{1}{2}\int_0^t h_s^\top C^{-1}h_s,\mathrm{d}s \right\}. \notag
\end{equation}
The quadratic form in the exponent can be written explicitly as
\begin{equation}
h_s^\top C^{-1}h_s = \frac{(h_s^I)^2-2\rho h_s^Ih_s^S+(h_s^S)^2}{1-\rho^2}. \notag
\end{equation}
Throughout the analysis, $\mathcal{H}$ is restricted to distortions that are locally square-integrable under this quadratic form, so that $\int_0^t h_s^\top C^{-1}h_s,\mathrm{d}s<\infty$, a.s. for every $t < \infty$. We further assume that $\xi^h$ is a true martingale, for instance under Novikov's condition or an equivalent standing condition. The process $\xi^h$ therefore defines an equivalent probability measure $\mathbb{Q}^h$ on each finite time interval.

By Girsanov’s theorem, there exists a subjective probability measure $\mathbb{Q}^h$ such that $\frac{\mathrm{d}\mathbb{Q}^h}{\mathrm{d} \mathbb{P}} \mid_{\mathcal{F}_t} = \xi_t^h$. Under this new measure, processes $W^{I,h} \triangleq \Big\{W^{I,h}_t: t \geq 0 \Big\}$ and $W^{S,h} \triangleq \Big\{W^{S,h}_t: t \geq 0 \Big\}$ defined by: 
\begin{align}
    \mathrm{d}W_t^{I,h} & = \mathrm{d}W^I_t - h_t^I\mathrm{d}t, \notag \\
    \mathrm{d}W_t^{S,h} & = \mathrm{d}W^S_t - h_t^S\mathrm{d}t,  \notag
\end{align}
are Brownian motions with instantaneous correlation $\rho$. Correspondingly, the individual reserve process \eqref{Benchmark Individual Wealth Process with Investment} evolves as: 
\begin{equation}
    \mathrm{d}m_t = \left[ x_t \eta \left( p_t + h^I_t \right) + y_t \sigma \left(q + h^S_t\right) + m_t r \right] \mathrm{d} t + x_t \eta \mathrm{d}W_t^{I, h} + y_t \sigma \mathrm{d}W_t^{S, h} + \mathrm{d}i_t - \mathrm{d}\delta_t. \notag
\end{equation} 
Accordingly, the aggregate capacity process for the entire insurance sector satisfies:  
\begin{equation}
    \mathrm{d}M_t = \left[ X_t \eta \left( p_t + h^I_t \right) + Y_t \sigma \left(q + h^S_t\right) + M_t r \right] \mathrm{d} t + X_t \eta \mathrm{d}W_t^{I, h} + Y_t \sigma \mathrm{d}W_t^{S, h} + \mathrm{d}I_t - \mathrm{d}\Delta_t. \notag
\end{equation} 
While in principle, the distortion processes $h^j$ may vary across insurers, we focus on a symmetric equilibrium in which all insurers behave identically. Thus, it is assumed that the optimally chosen $h^j$ takes the same value across all $j \in \mathcal{J}$. 

With model uncertainty, each insurer’s optimization problem deviates from \eqref{Benchmark Optimization Objective} and is evaluated under the subjective measure $\mathbb{Q}^h$, incorporating a penalty term that reflects the insurer’s preference for robustness against model misspecification. Following the variational preferences framework of \citet{anderson2003quartet}, \citet{maccheroni2006ambiguity}, and \citet{maccheroni2006dynamic}, the insurer’s objective can be written as: 
\begin{equation}
    \inf_{\mathbb{Q}^h \in \mathcal{Q}} \mathbb{E}^{\mathbb{Q}^h} \left\{ \int_0^{\infty} e^{-\lambda t} \Big[ \mathrm{d}\delta_t - (1 + \gamma) \mathrm{d} i_t \Big] \right\} + \mathcal{K}(\mathbb{Q}^h), \label{Robust Optimization Objective}
\end{equation}
where $\mathcal{Q}$ denotes the set of admissible probability measures $\mathbb{Q}^h$, and $\mathcal{K}(\mathbb{Q}^h)$ captures the entropy cost of model distortion: 
\begin{equation}
    \mathcal{K}(\mathbb{Q}^h) = \frac{1}{2} \mathbb{E}^{\mathbb{Q}^h} \left[ \int_0^\infty e^{-\lambda t} \Theta_t \frac{(h_t^I)^2 - 2\rho h_t^I h_t^S + (h_t^S)^2}{1-\rho^2} \mathrm{d}t \right]. \notag
\end{equation}
This term represents a discounted, state-weighted relative entropy between $\mathbb{Q}^h$ and the reference model $\mathbb{P}$, where $\Theta_t$ determines the sensitivity to model perturbations (i.e., the cost of distortion). The components $h^I$ and $h^S$ are the worst-case drift distortions for the insurance and financial risk factors, respectively. The insurer’s value function, denoted by $v(m, M)$, depends jointly on the individual liquid reserves $m$ and the aggregate capacity $M$. In the following analysis, we aim to eliminate one of these state variables to simplify the characterization of equilibrium. 

For tractability, we assume that $\Theta_t(m_t) = \theta m_t$, where $\theta > 0$ measures the overall degree of concern for robustness.\footnote{Equivalently, $1/\theta$ captures the degree of ambiguity aversion \citep{maccheroni2006ambiguity, maccheroni2006dynamic}, such that a larger $\theta$ corresponds to lower aversion to model uncertainty.} It yields two useful implications. First, the entropy cost is proportional to the insurer’s liquid reserves, which is economically intuitive since larger insurers incur greater distortion costs. Second, it preserves the homogeneity of the optimization problem \eqref{Robust Optimization Objective} in $(m, x, \mathrm{d}\delta, \mathrm{d}i)$, implying that the value function must also be homogeneous of degree one in $m$. Accordingly, we posit that the value function takes the form $v(m, M) = m u(M)$, where $u(M)$ represents the market-to-book ratio of each insurer, identical across the industry in equilibrium. Conceptually, $u(M)$ can be interpreted as the insurance analogue of Tobin’s $q$ ratio.

\section{Equilibrium Solution} \label{Section Equilibrium}

In line with \citet{henriet2016dynamics}, as well as with many related liquidity management problems, we focus on a barrier-type candidate solution. Specifically, let $\underline{M}$ and $\overline{M}$ denote the external financing and payout boundaries, respectively, whose values are to be determined endogenously by the equilibrium conditions. The following regularity conditions are imposed on any candidate solution and are later checked numerically. 

\begin{Assumption} \label{Assumption market-to-book ratio}
    We impose the following conditions on the equilibrium outcomes, which can be numerically verified under reasonable parameter settings in Section \ref{Section Numerical}: 
    
    (1) Both $v(\cdot, \cdot)$ and $u(\cdot)$ are smooth functions, continuously differentiable up to the second order. 
    
    (2) For $M \in [\underline{M}, \overline{M}]$, it holds that $u(M) \geq 0$ and $u^{\prime}(M) \leq 0$.  In other words, the market-to-book ratio of the insurance sector is assumed to be a decreasing function of aggregate capacity. 

    (3) For the external financing and payout boundaries, we impose $0< \underline{M} < \overline{M}$. In addition, $m \geq 0$ is required, as it would be unrealistic to assume that either an individual insurer or the entire industry continues underwriting when liquid reserves are negative.  
\end{Assumption}

This assumption would simplify the equilibrium analysis and ensure that the equilibrium possesses well-behaved properties. The monotonicity of $u(\cdot)$ is crucial for guaranteeing that the optimal policy takes a barrier-type form. Otherwise, the payout and recapitalization regions could become disconnected, leading to multiple thresholds and more complex strategic behavior. The strict positivity of the lower boundary is imposed at this stage and justified below, where we show that $\underline{M}=0$ is incompatible with the verification system. 

In the presence of model uncertainty, the insurer’s optimization problem \eqref{Robust Optimization Objective} becomes a robust control problem. Based on the dynamic programming principle, the insurer’s value function $v(m, M)$ is characterized by the following Hamilton-Jacobi-Bellman-Isaacs (HJBI) equation: 
\begin{align}
    \lambda v(m ,M) = & \sup_{x \geq 0, y, \mathrm{d}\delta \geq 0, \mathrm{d}i \geq 0} \inf_{h \in \mathcal{H}} \bigg\{ \frac{1}{2} \Theta(m) \frac{(h^I)^2 - 2\rho h^Ih^S + (h^S)^2}{1-\rho^2} + \left[ x \eta (p + h^I) + y \sigma (q + h^S) + mr \right] v_m \notag \\
    & + \left[ D(p) \eta (p + h^I) + Y \sigma (q + h^S) + Mr \right] v_M + \frac{1}{2} \left( x^2 \eta^2 + 2 \rho x \eta y \sigma + y^2\sigma^2 \right) v_{mm} \notag \\
    & + \frac{1}{2} \left( D(p)^2 \eta^2 + 2 \rho D(p) \eta Y \sigma  + Y^2 \sigma^2 \right) v_{MM} + \left(x D(p) \eta^2 + y Y \sigma^2 + \rho x \eta Y \sigma + \rho D(p) \eta y \sigma  \right) v_{mM} \notag \\
    & + \mathrm{d}\delta (1 - v_m) + \mathrm{d}i (v_m - 1 - \gamma)\bigg\}. \label{HJBI Original}
\end{align}
Equivalently, if $\mathcal{A}v$ denotes the regular-control Hamiltonian obtained from \eqref{HJBI Original} after setting the singular controls to zero and subtracting $\lambda v$, then the singular-control problem has the variational-inequality form
\begin{equation}
    \max\left\{\mathcal{A}v,\;1-v_m,\;v_m-1-\gamma\right\}=0. \notag
\end{equation}
The dividend term corresponds to reflection at the payout boundary where $v_m=1$, while the recapitalization term corresponds to reflection at the financing boundary where $v_m=1+\gamma$. In the internal financing region both gradient constraints are slack and the regular HJBI equation holds with equality.

\subsection{Internal Financing Region} 

We begin by analyzing the internal financing region, where the optimal dividend and recapitalization policies satisfy $\mathrm{d}\delta^{\ast} = 0$ and $\mathrm{d}i^{\ast} = 0$. For $m = 0$, we have $\Theta(m) = 0$, and the minimization with respect to $h^I$ and $h^S$ admits an interior solution if and only if: 
\begin{equation}
    x^{\ast}(0, M) = - \left. \frac{m u^{\prime}(M)}{u(M)} \right|_{m = 0} D(p) = 0, \quad y^{\ast}(0, M) = - \left. \frac{m u^{\prime}(M)}{u(M)} \right|_{m = 0} Y = 0, \notag
\end{equation} 
given the assumption that $v(m, M) = m u(M)$. Then, the HJBI equation holds trivially, as both sides are equal to zero. This result implies that an individual insurer does not engage in underwriting or investment activities when its liquid reserves are depleted. 

For the more general case $m>0$, the terms in the HJBI equation that depend on the drift distortion $h=(h^I,h^S)^\top$ are
\begin{equation}
\inf_h \left\{ \frac12\Theta(m)h^\top C^{-1}h +\eta\big(xv_m+D(p)v_M\big)h^I +\sigma\big(yv_m+Yv_M\big)h^S \right\}. \notag
\end{equation}
The first-order condition is therefore
\begin{equation}
\Theta(m)C^{-1}h
+
\begin{pmatrix}
\eta\big(xv_m+D(p)v_M\big) \\
\sigma\big(yv_m+Yv_M\big)
\end{pmatrix}
=0.
\notag
\end{equation}
Solving this linear system gives the worst-case drift distortion 
\begin{align}
    {h^{I}} & = - \frac{\eta\left(x v_m + D(p) v_M\right)+\rho\sigma (y v_m + Y v_M)}{\Theta(m)}, \label{h1 candidate} \\
    {h^{S}} & = - \frac{\rho\eta\left(x v_m + D(p) v_M\right)+\sigma (y v_m + Y v_M)}{\Theta(m)}. \label{h2 candidate}
\end{align}
Substituting the minimizing distortion into the HJBI and applying the envelope theorem, the first-order condition with respect to $x$ can be simplified under $v(m,M)=m u(M)$ to
\begin{equation}
    h^I = - \frac{u^{\prime}(M)}{u(M)} \Big( D(p) \eta + \rho Y \sigma \Big) - p. \label{hi solution}
\end{equation}
Similarly, the first-order condition with respect to $y$ is equivalent to
\begin{equation}
    h^S = -\frac{u^{\prime}(M)}{u(M)} \Big( Y \sigma + \rho D(p) \eta \Big) - q. \label{hs solution}
\end{equation}
Combining equations \eqref{h1 candidate}, \eqref{h2 candidate}, \eqref{hi solution} and \eqref{hs solution}, we obtain: 
\begin{align}
    x^{\ast}(m, M) & = m \left\{ \frac{\theta}{u(M) \eta} \Big[ \frac{p - \rho q}{1 - \rho^2} - R(M) D(p) \eta \Big] + R(M) D(p)  \right\}, \label{Optimal x} \\
    y^{\ast}(m, M) & = m \left\{ \frac{\theta}{u(M) \sigma} \Big[ \frac{q - \rho p}{1 - \rho^2} - R(M) Y \sigma \Big] + R(M) Y  \right\}, \label{Optimal y}
\end{align}
where $R(M) \triangleq - \frac{u^{\prime}(M)}{u(M)}$ denotes the market risk-aversion, which simplifies the expressions above. 

Here, the candidate underwriting scale $x^{\ast}(m, M)$ is optimal whenever it is non-negative. Otherwise, due to the non-negativity constraint, the optimal scale is restricted to $x^{\ast}(m, M) = 0$. While our main interest lies in the non-binding case, where the insurance market operates normally without persistent zero-underwriting equilibria, we additionally impose the following assumption. 

\begin{Assumption} \label{Assumption Robust}
    We impose the following conditions on parameters and equilibrium outcomes: 

    (1) Parameters of the insurance and financial markets should satisfy $\left|\rho\right| < \frac{\alpha \eta}{q}$. 

    (2) For $M \in [\underline{M}, \overline{M}]$, the market-to-book ratio function satisfies $M R(M) < 1$. 
\end{Assumption}

The first condition indicates that the correlation between the insurance and financial market risks cannot be excessively high. The second condition implies that the implicit market risk aversion cannot be too large and is bounded by the inverse of the aggregate capacity. Again, it can be verified numerically in Section~\ref{Section Numerical}. Given Assumptions~\ref{Assumption market-to-book ratio} and~\ref{Assumption Robust}, we can further pre-assume that $f(M) \triangleq \frac{\theta}{u(M) \eta} \Big[ \frac{p - \rho q}{1 - \rho^2} - R(M) D(p) \eta \Big] + R(M) D(p) > 0$, so that $x^{\ast}(m, M)$ given in \eqref{Optimal x} indeed represents the optimal control.\footnote{Suppose for some $M$, $f(M) \leq 0$. Then $x^{\ast}(m, M) = 0$ for all $m$. In this case, $D(p^{\ast}(M)) = 0$ and $p^{\ast}(M) = \alpha \eta$. Consequently, $f^{\ast}(M) = \frac{\theta}{u^{\ast}(M) \eta} \frac{\alpha \eta - \rho q}{1 - \rho^2} > 0$, given Assumptions \ref{Assumption market-to-book ratio} and \ref{Assumption Robust}, which leads to a contradiction. } 

Based on equations~\eqref{Optimal x} and~\eqref{Optimal y}, the market-clearing condition for the insurance market and the equilibrium condition for financial investment are given by: 
\begin{align}
    D(p^{\ast}(M)) & = M \left\{ \frac{\theta}{u(M) \eta} \Big[ \frac{p - \rho q}{1 - \rho^2} - R(M) D(p(M)) \eta \Big] + R(M) D(p(M))  \right\}, \notag \\
    Y^{\ast}(M) & = M \left\{ \frac{\theta}{u(M) \sigma} \Big[ \frac{q - \rho p}{1 - \rho^2} - R(M) Y(M) \sigma \Big] + R(M) Y(M)  \right\},  \notag
\end{align}
which represent the aggregate underwriting and investment activities across all insurers, satisfying $\int_{\mathcal{J}} m^j \mathrm{d}j = M$. Solving these conditions yields:
\begin{align}
    D(p^{\ast}(M)) & = \frac{p^*(M)-\rho q}{A(M)(1-\rho^2) \eta}, \notag \\
    Y^{\ast}(M) & =  \frac{q-\rho p^*(M)}{A(M)(1-\rho^2) \sigma}, \notag
\end{align}
where $A(M) \triangleq R(M)+\frac{u(M)}{\theta}\left(\frac{1}{M}-R(M)\right)$. Although $A(M)$ contains the term $\frac{1}{M}$, it is only defined on the invariant state space $[\underline M,\overline M]$ with $\underline M>0$. 

Given the demand function $D(p)=1-\frac{p}{\alpha\eta}$, the closed-form equilibrium price is
\begin{equation}
    p^*(M) = \frac{A(M)(1-\rho^2)\eta+\rho q}{1 + \frac{1}{\alpha} A(M)(1-\rho^2)}. \label{express p by u}
\end{equation}
And the corresponding equilibrium demand is
\begin{equation}
    D(p^*(M)) = \frac{1 - \frac{\rho q}{\alpha \eta}}{1 + \frac{1}{\alpha} A(M) (1 - \rho^2) }.  \label{express dp by u}
\end{equation}
Here, it can be verified that $D(p^*(M)) > 0$ due to the assumption $\left|\rho\right| < \frac{\alpha \eta}{q}$. And the equilibrium aggregate investment is expressed as 
\begin{equation}
    Y^*(M)= \frac{q + A(M) \left( \frac{q}{\alpha} - \rho \eta \right)}{A(M) \sigma \left[1 + \frac{1}{\alpha} A(M) (1 - \rho^2) \right]}. \label{Y solution}
\end{equation}

Hence, all equilibrium quantities can be fully characterized as functions of the market-to-book ratio $u(\cdot)$. Finally, substituting these expressions into the HJBI equation yields
\begin{align}
\lambda-r ={} & \frac{\theta}{2u(M)}
\frac{ \big(h^{I,*}(M)\big)^2 -2\rho h^{I,*}(M)h^{S,*}(M) +\big(h^{S,*}(M)\big)^2}{1-\rho^2} -MrR(M) \notag \\
& + \left( \frac{1}{2}\frac{u''(M)}{u(M)} -R(M)^2 \right) \Big( D(p^*(M))^2\eta^2 +2\rho D(p^*(M))\eta Y^*(M)\sigma +Y^*(M)^2\sigma^2 \Big), \label{ODE to u}
\end{align}
which reduces to an ordinary differential equation (ODE) in $u(\cdot)$ that can be solved numerically.

\subsection{External Financing and Payout Regions} 

We now turn to the analysis of optimal dividend and recapitalization strategies. Based on the HJBI equation~\eqref{HJBI Original}, the maximization with respect to $\mathrm{d}\delta$ and $\mathrm{d}i$ is independent of the choice of $h$. A necessary condition for a candidate verification solution is: 
\begin{equation}
    1 - v_m \leq 0, \quad \text{and} \quad  v_m - 1 - \gamma \leq 0, \notag
\end{equation}
which implies that $1 \leq u(M) \leq 1 + \gamma$. Moreover, the optimal payout policy satisfies $\mathrm{d}\delta^{\ast} > 0$ only when $u(M) = 1$, whereas the optimal recapitalization policy satisfies $\mathrm{d}i^{\ast} > 0$ only when $u(M) = 1 + \gamma$. Consequently, the value-matching conditions at the two boundaries are given by:  
\begin{equation}
    u(\underline{M}) = 1 + \gamma, \quad u(\overline{M}) = 1. \notag
\end{equation}

Referring to \citet{henriet2016dynamics} and \citet{luciano2022fluctuations}, we further impose a no-arbitrage condition. Let $V(M) = M u(M)$ denote the aggregate market value of the insurance industry. The absence of arbitrage opportunities requires the following conditions:
\begin{align}
    & V^{\prime}(\overline{M}) = u(\overline{M}) + \overline{M} u^{\prime}(\overline{M}) = 1 \quad \Longleftrightarrow \quad u^{\prime}(\overline{M}) = 0,  \notag \\
    & V^{\prime}(\underline{M}) = u(\underline{M}) + \underline{M} u^{\prime}(\underline{M}) = 1 + \gamma \quad \Longleftrightarrow \quad u^{\prime}(\underline{M}) = 0 \quad \text{or} \quad \underline{M} = 0. \notag
\end{align}
Economically, these conditions ensure that the marginal change in the aggregate market value induced by a dividend payout or a recapitalization equals the marginal flow of funds withdrawn from or injected into the industry by shareholders. 

Here, we prove that the contingent condition $\underline{M} = 0$ cannot generally hold in our setting. Suppose $\underline{M} = 0$ and there exists a solution $\left(p^{\ast}(M), u^{\ast}(M)\right)$ for $M \in [0, \overline{M}]$. Then, the market-clearing conditions imply that $D(p^{\ast}(0)) = 0$, $p^{\ast}(0) = \alpha \eta$, and $Y^{\ast}(0) = 0$. Since equation~\eqref{ODE to u} holds for $M > 0$, it should also hold at $M = 0$, provided that $u(\cdot)$ is sufficiently smooth and bounded in a neighborhood of $M = 0$. Substituting $\underline{M} = 0$ and $D(p^{\ast}(0)) = 0$ into the equation yields: 
\begin{equation}
    \lambda - r = \frac{\theta}{2 (1 + \gamma)} \frac{\alpha^2 \eta^2 - 2 \rho \alpha \eta q + q^2 }{1 - \rho^2}, \notag
\end{equation} 
which does not generally hold unless specific parameter restrictions are imposed. Therefore, the alternative boundary condition $u^{\prime}(\underline{M}) = 0$ with $\underline{M} > 0$ must apply.  

\subsection{Robust Equilibrium}

Combining the verification solution to the control problem with the determination of the external financing and payout boundaries, we characterize the candidate market equilibrium in the presence of model uncertainty as follows. 

\begin{Proposition}[\textbf{Robust Equilibrium with Model Uncertainty}] \label{Proposition Robust Equilibrium}
    In the presence of model uncertainty, suppose that Assumptions \ref{Assumption market-to-book ratio} and \ref{Assumption Robust} hold and that the following system of equations admits a solution for $p^{\ast}(M)$ and $u^{\ast}(M)$ on $[\underline{M}, \overline{M}]$:  
    \begin{equation}
        \left\{
        \begin{aligned}
            & p(M) = \frac{\left[R(M)+\frac{u(M)}{\theta}\left(\frac{1}{M}-R(M)\right)\right](1-\rho^2)\eta+\rho q}{1 + \frac{1}{\alpha}\left[R(M)+\frac{u(M)}{\theta}\left(\frac{1}{M}-R(M)\right)\right](1-\rho^2)}, \\
            & \lambda - r =  \frac{\theta}{2u(M)} \frac{ \big(h^{I,*}(M)\big)^2 -2\rho h^{I,*}(M)h^{S,*}(M) +\big(h^{S,*}(M)\big)^2}{1-\rho^2} -MrR(M)  \\
            & + \left( \frac{1}{2}\frac{u''(M)}{u(M)} -R(M)^2 \right) \Big( D(p^*(M))^2\eta^2 +2\rho D(p^*(M))\eta Y^*(M)\sigma +Y^*(M)^2\sigma^2 \Big),  
        \end{aligned}
        \right. \label{Solution to Robust Equilibrium}
    \end{equation}
    subject to the boundary conditions: $u(\underline{M}) = 1 + \gamma$, $u(\overline{M}) = 1$, $u^{\prime}(\underline{M}) = u^{\prime}(\overline{M}) = 0$. Then, there exists a stationary Markovian equilibrium characterized as follows: 

    (1) For $M \in (\underline{M}, \overline{M})$, the market price of insurance is $p^{\ast}(M)$, each insurer's market-to-book value function is $u^{\ast}(M)$, the worst-case drift distortion is $h^{\ast}(M) = (h^{I,*}(M), h^{S,*}(M))^\top$ as given in \eqref{hi solution} and \eqref{hs solution}, and the aggregate investment amount is $Y^{\ast}(M)$ as given in \eqref{Y solution}. For an insurer with reserves $m$, the optimal underwriting amount is $x^{\ast}(m, M) = \frac{m}{M} D(p^{\ast}(M))$ as given in \eqref{Optimal x}, the optimal investment amount is $y^{\ast}(m, M) = \frac{m}{M} Y^{\ast}(M)$ as given in \eqref{Optimal y}, and the shareholders’ value is $v^{\ast}(m, M) = m u^{\ast}(M)$. 
    
    (2) For $M \geq \overline{M}$, insurers distribute dividends to shareholders until the aggregate liquid reserves fall below $\overline{M}$. For $M \leq \underline{M}$, insurers raise external capital to restore reserves to the level of $\underline{M}$. 
\end{Proposition} 

While the existence and uniqueness of equilibrium have been established in previous studies such as \citet{henriet2016dynamics}, we do not attempt to provide a rigorous proof here, as the introduction of model uncertainty substantially complicates the analytical structure of the solution. Instead, our numerical analysis in Section~\ref{Section Numerical} demonstrates that, under standard parameter specifications, the ODE~\eqref{ODE to u} indeed admits a solution that satisfies Assumptions~\ref{Assumption market-to-book ratio} and~\ref{Assumption Robust}. 

\citet{pang2026robust} investigates a similar robust equilibrium in an insurance market without outside financial investment opportunities. If we set $\rho = 0$, $r = 0$ and $q = 0$, the system of equations in~\eqref{Solution to Robust Equilibrium} collapses to the results in \citet{pang2026robust}. In contrast, introducing either a non-zero correlation $(\rho \neq 0)$ or a non-zero financial return $(q \neq 0)$ leads to qualitatively different equilibrium outcomes. 

\begin{Proposition}[\textbf{Correlation and Equilibrium Loading}] \label{Proposition sign of loading}
Given the existence of the equilibrium in Proposition \ref{Proposition Robust Equilibrium}, we have $p^*(M)>0$ for all $M\in[\underline{M},\overline{M}]$ whenever $\rho\geq0$. If $\rho<0$, then $p^*(M) < 0$ at state $M$ if and only if 
\begin{equation}
-\rho q > A(M)(1-\rho^2)\eta. \notag
\end{equation}
\end{Proposition}

In the classical insurance economics literature, the insurance loading is often imposed as an exogenous positive constant. This convention is analytically convenient but makes it difficult to explain why underwriting losses are frequently observed in practice. For example, more than half of Chinese property-casualty insurers have reported underwriting losses for several consecutive years. \citet{chen2025dynamic} propose an investment-driven pricing mechanism to rationalize this pattern. A similar mechanism operates in our framework. When underwriting surplus and financial investment returns are negatively correlated, investment gains can hedge underwriting losses. In a competitive market, this hedging benefit is passed through to policyholders through lower premiums and, when the hedge is sufficiently strong, negative loadings.

\section{Quantitative Analysis} \label{Section Numerical}

In this section, we perform numerical simulations to solve the ODE system and characterize the dynamic equilibrium of the insurance market. As the benchmark setting, we specify the parameter values as follows. The discount rate is set to $\lambda = 4\%$ for shareholders’ valuation of future cash flows. Following the estimation of \citet{luciano2022fluctuations}, we set the unit loss to $l = 1.0$ and the volatility of unit loss risk to $\eta = 28\%$. The risk-free rate is $r = 1.528\%$. The portfolio of risky assets consists of 80\% bonds and 20\% equity stakes, implying a weighted-average expected excess return of $\mu - r = 2.0\%$ and a volatility of $\sigma = 18\%$. The cost of external financing is parameterized as $\gamma = 20\%$, interpreted as the expected return on equity financing, while the risk aversion coefficient of insurees (general households) is set to $\alpha = 2.0$. For the correlation between the insurance and financial markets, due to the limited empirical evidence in the literature, we assume a moderate negative value of $\rho = -0.4$. This assumption is broadly consistent with recent macro-financial conditions, where slower economic growth has been accompanied by higher insurance payouts and lower interest rates, which in turn increase the value of fixed-income investments. Finally, following \citet{ling2021robust}, we adopt $\theta = 2.8$ as the benchmark degree of robustness concern. 

In the numerical computation, the free boundaries are solved jointly with the state variables, starting from a smooth initial guess until convergence is achieved. The equilibrium outcomes are found to satisfy Assumptions~\ref{Assumption market-to-book ratio} and~\ref{Assumption Robust}, confirming the internal consistency of the numerical solution.  

\begin{table}[htbp]
\centering
\setlength{\tabcolsep}{20pt}
\renewcommand{\arraystretch}{1.5}
\small
\begin{tabular}{lcc}
\hline
Parameter & Symbol & Value \\
\hline
Discount factor                & $\lambda$       & 4\% \\
Expected losses on insurance risk       & $l$    & 1.0 \\
Volatility of losses on insurance risk       & $\eta$    & 28\% \\
Risk-free rate                & $r$    & 1.528\% \\
Expected return on risky assets                & $\mu$    & 3.528\% \\
Volatility of risky assets               & $\sigma$    & 18\% \\
Cost of external financing    & $\gamma$  & 20\% \\
Risk aversion of insuree      & $\alpha$  & 2.0 \\
Robustness degree          & $\theta$  & 2.8 \\
Correlation coefficient         & $\rho$  & -0.4 \\
\hline
\end{tabular}
\caption{Benchmark Parameter Values.}
\end{table}

\subsection{Comparison of Equilibria with and without Financial Investment} \label{Subsection Comparison of Equilibria}

\begin{figure}[tbp]
    \centering
    \includegraphics[width=0.45\linewidth]{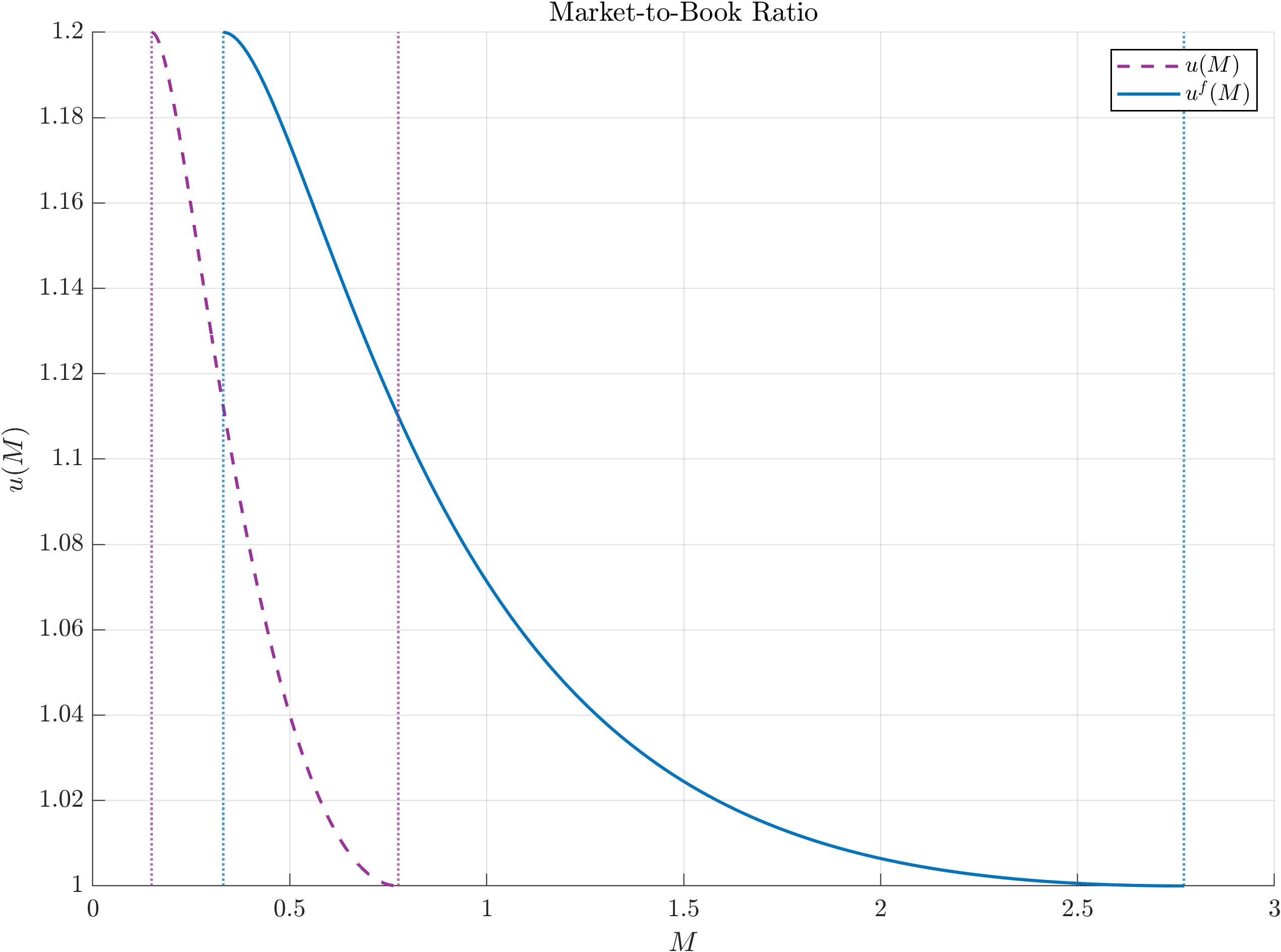}
    \includegraphics[width=0.45\linewidth]{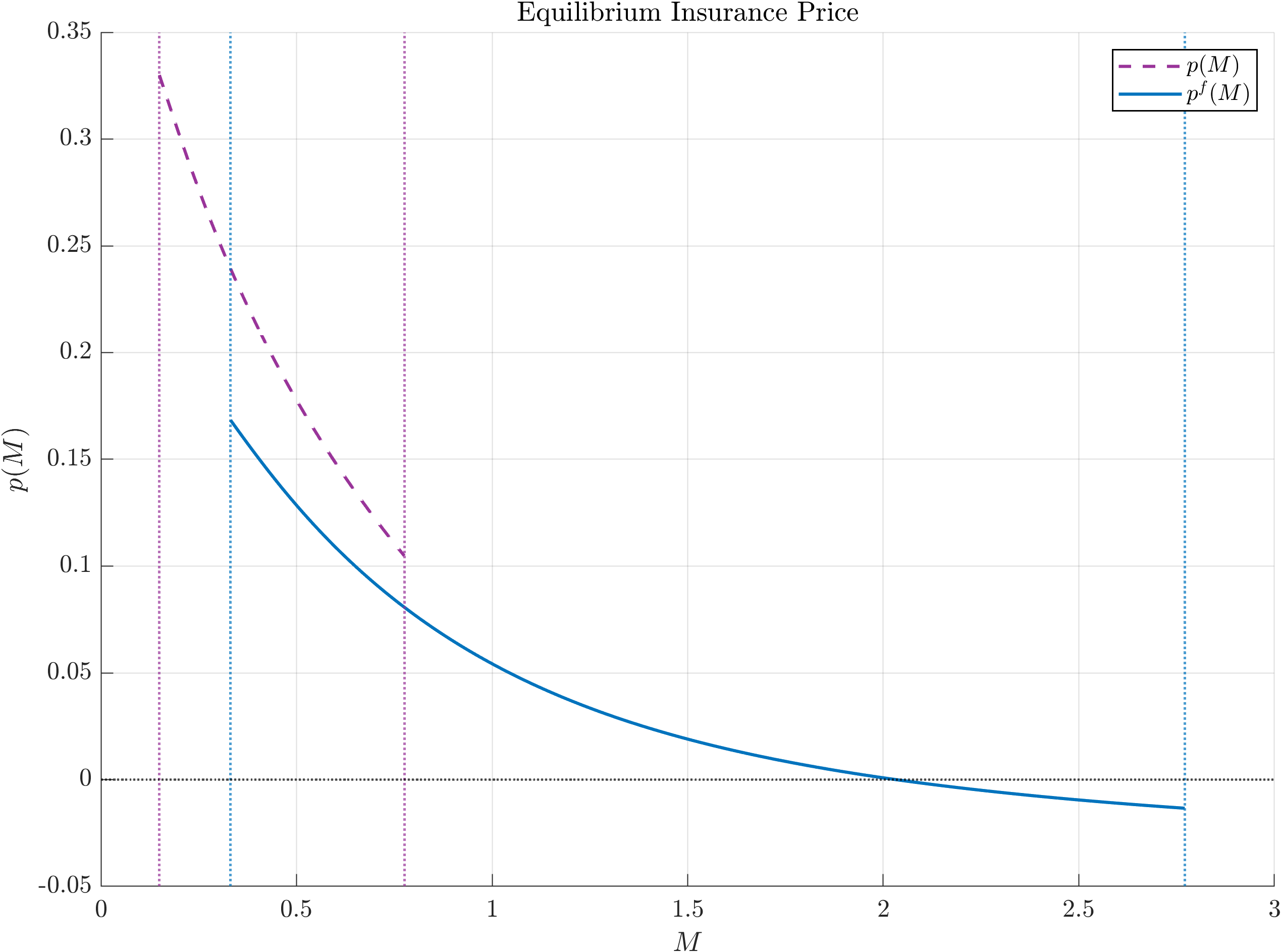}
    \includegraphics[width=0.45\linewidth]{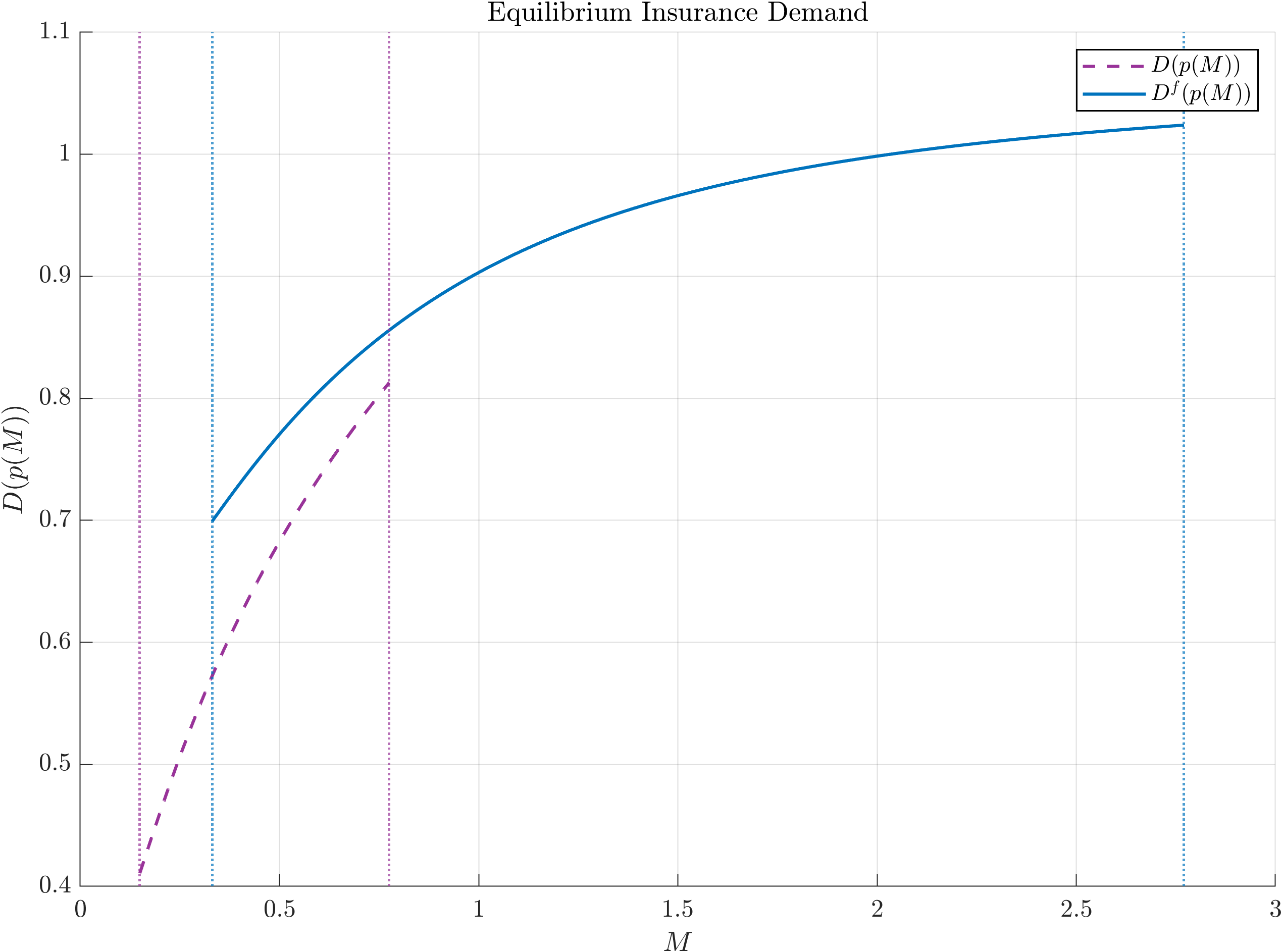}
    \includegraphics[width=0.45\linewidth]{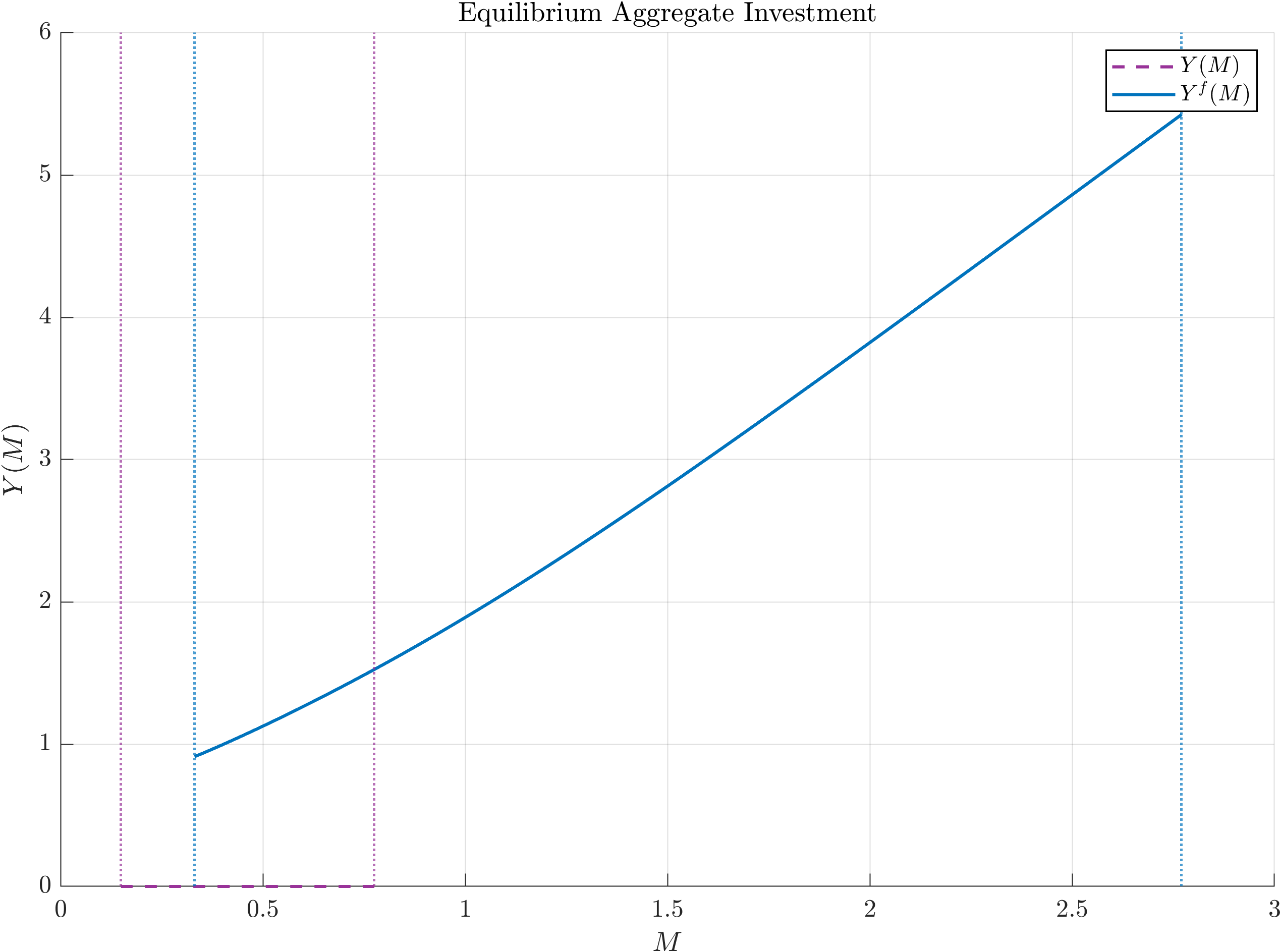}    
    \caption{Equilibrium Outcomes with and without Financial Investment.}
    \label{Figure Comparison of Equilibrium Outcomes}
\end{figure}

Figure~\ref{Figure Comparison of Equilibrium Outcomes} reports the numerically solved market-to-book ratio and equilibrium insurance price under the benchmark parameter setting (blue solid curves), and compares them with the results of \citet{pang2026robust}, which correspond to the case without financial investment opportunities (i.e., $\rho = 0$, $r = 0$, and $q = 0$, shown as purple dashed curves). Several notable features can be observed. 

First, the minimum, maximum, and range of aggregate capacity all expand when accounting for outside financial investment. Without financial investment, $\underline{M} = 0.1494$, $\overline{M} = 0.7755$, and $\Delta M \triangleq \overline{M} - \underline{M} = 0.6261$, while with investment $\underline{M}^f = 0.3312$, $\overline{M}^f = 2.7705$, and $\Delta M^f = 2.4394$. The expansion of both boundaries reflects that the introduction of financial investment opportunities alters insurers' liquidity management behavior. The higher lower boundary $\underline{M}^f$ indicates that insurers require a larger minimum reserve to buffer potential losses from financial market volatility and correlated risks, implying tighter recapitalization conditions. Meanwhile, the higher upper boundary $\overline{M}^f$ results from the accumulation of investment returns, which allows insurers to build larger reserves before distributing dividends. Overall, these adjustments widen the equilibrium range of aggregate capacity, indicating that the insurance market can sustain greater fluctuations. 

Second, consistent with the verification conditions, the numerically solved market-to-book ratio function is decreasing in aggregate capacity. At the same capacity level, the market-to-book ratio in the robust equilibrium with financial investment is significantly higher. This reflects that the presence of financial investment opportunities enhances insurers' franchise value by providing an additional channel for return generation beyond underwriting activities. The improved investment income strengthens insurers' balance sheets and raises shareholders' valuation of the firm. Moreover, as the volatility and correlation of financial assets are partially diversified through equilibrium portfolio allocation, insurers' exposure to aggregate risk becomes more manageable, further increasing their market valuation relative to book equity. 

Third, consistent with \citet{henriet2016dynamics} and \citet{luciano2022fluctuations}, the equilibrium insurance price decreases with aggregate capacity, in line with the empirical observation that insurers tend to lower premiums when capital is abundant. Compared with \citet{pang2026robust}, the overall level of equilibrium prices differs because financial investment changes both aggregate risk exposure and the ambiguity penalty. Under the benchmark value $\rho=-0.4$, the price crosses zero near the payout boundary, with $\min_M p^\ast(M)=-0.0134$. This pattern is consistent with the investment hedging channel in Proposition~\ref{Proposition sign of loading}. When underwriting surplus and financial returns are negatively correlated, investment gains can hedge underwriting losses. In a competitive market, part of this hedging value is passed through to policyholders through lower premiums, so negative loadings can arise even in a robust pricing framework. 

\begin{figure}[htbp]
    \centering
    \includegraphics[width=0.45\linewidth]{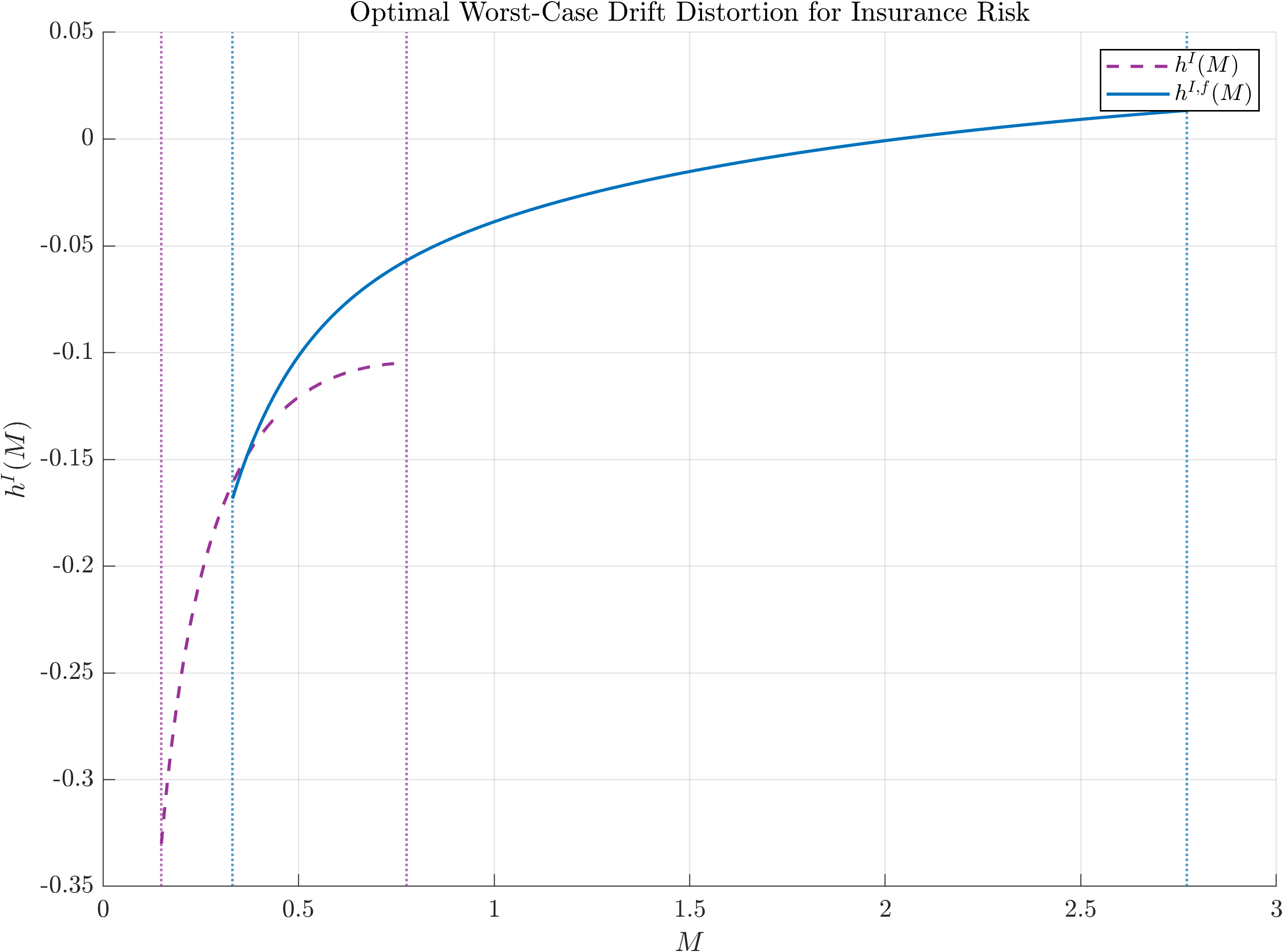}
    \includegraphics[width=0.45\linewidth]{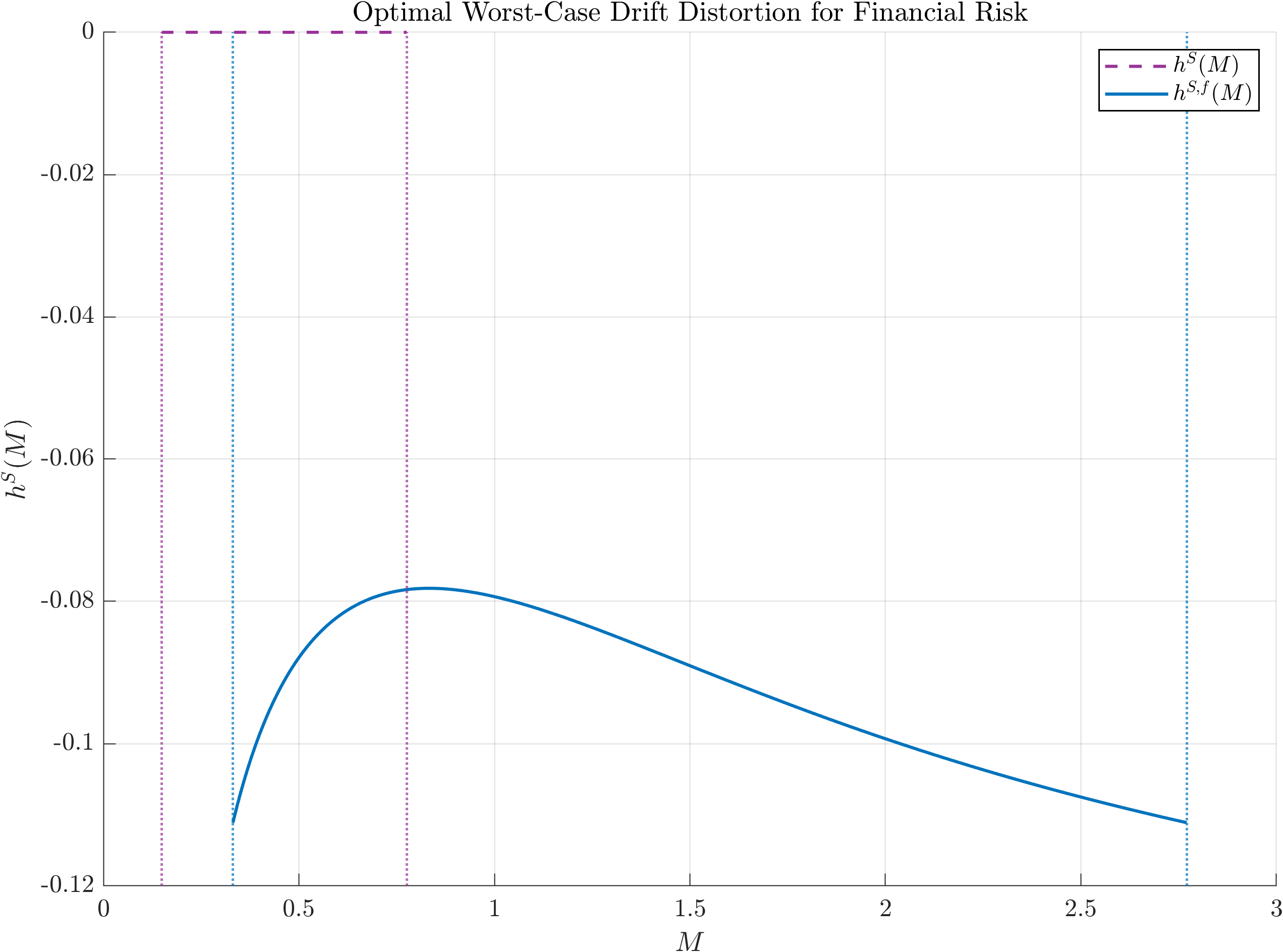}
    \caption{Optimal Worst-Case Drift Distortions with and without Financial Investment.}
    \label{Figure Worst-Case Drift Distortion}
\end{figure} 

Figure~\ref{Figure Worst-Case Drift Distortion} illustrates the optimal worst-case drift distortions associated with insurance and financial risks. These distortions show how insurers adjust the perceived drifts of the correlated Brownian risk factors under model uncertainty. While $p^{\ast}(M)$ represents the market price of insurance risk, $h^{I,\ast}(M)$ captures the insurance-risk component of the ambiguity adjustment. Following the robust-control interpretation of \citet{anderson2003quartet}, this term can be viewed as the shadow price associated with potential model misspecification. Using the equilibrium conditions, we obtain
\begin{equation}
h^{I,\ast}(M)+p^\ast(M) = R(M)\frac{p^\ast(M)}{A(M)}, \qquad h^{S,\ast}(M)+q = R(M)\frac{q}{A(M)}. \notag
\end{equation}
Since $R(M)\geq0$ and $A(M)>0$, the drift-adjusted underwriting margin has the same sign as $p^\ast(M)$, whereas the drift-adjusted financial investment margin is non-negative whenever $q\geq0$. Thus, when $p^\ast(M)<0$, underwriting by itself has a negative margin even under the worst-case measure. The insurer is nevertheless compensated for its combined risky position, because
\begin{align}
& D(p^\ast(M))\eta\Big(p^\ast(M)+h^{I,\ast}(M)\Big) +Y^\ast(M)\sigma\Big(q+h^{S,\ast}(M)\Big) \notag \\
= {}& R(M) \Big( D(p^\ast(M))^2\eta^2 +2\rho D(p^\ast(M))\eta Y^\ast(M)\sigma +Y^\ast(M)^2\sigma^2 \Big) \geq0. \notag
\end{align}
Negative loading is therefore a component-level pricing outcome generated by the investment hedge, rather than a violation of overall risk compensation.

\subsection{Impact of Correlation Coefficient} 

\begin{figure}[t]
    \centering
    \includegraphics[width=0.45\linewidth]{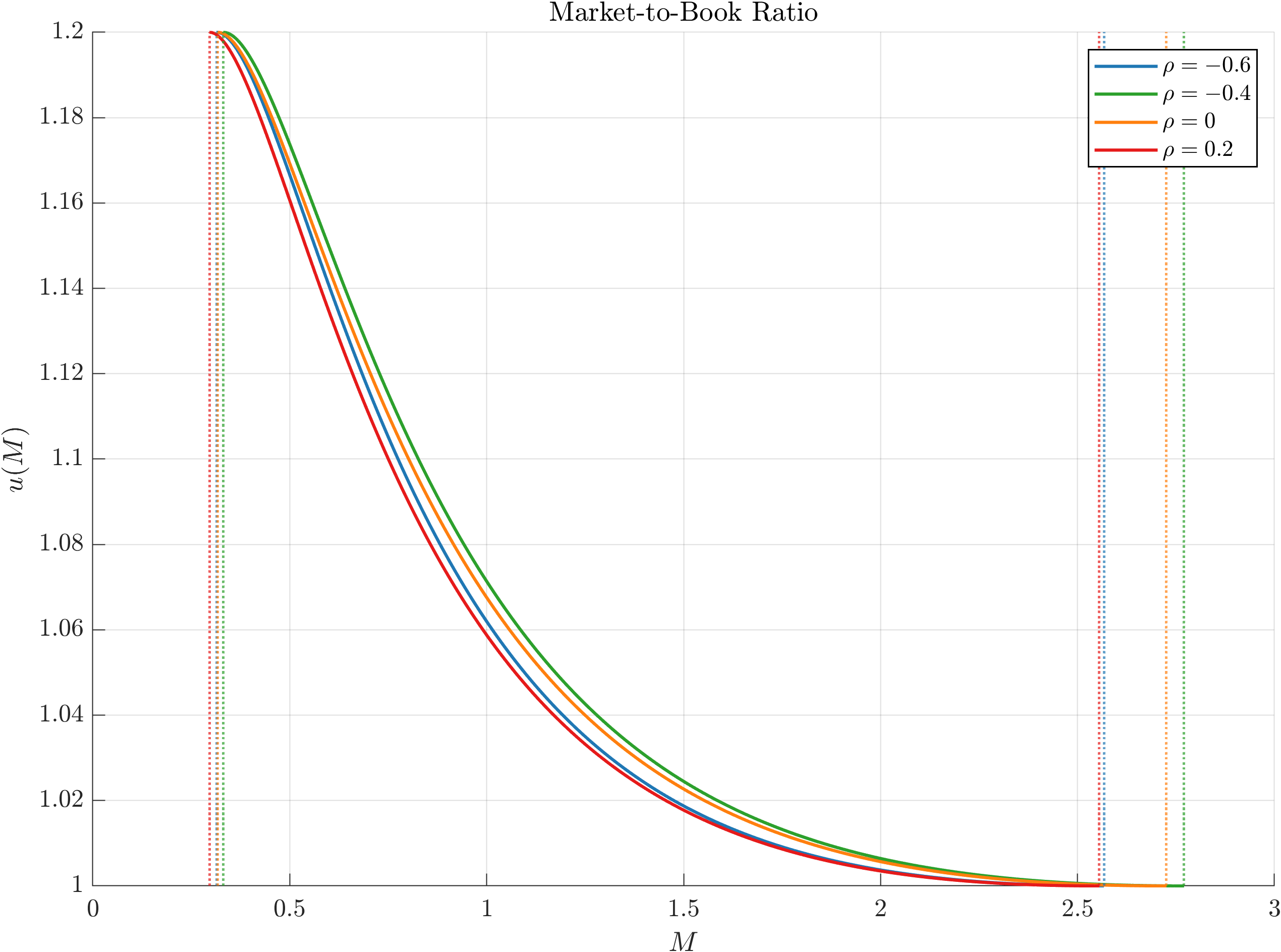}
    \includegraphics[width=0.45\linewidth]{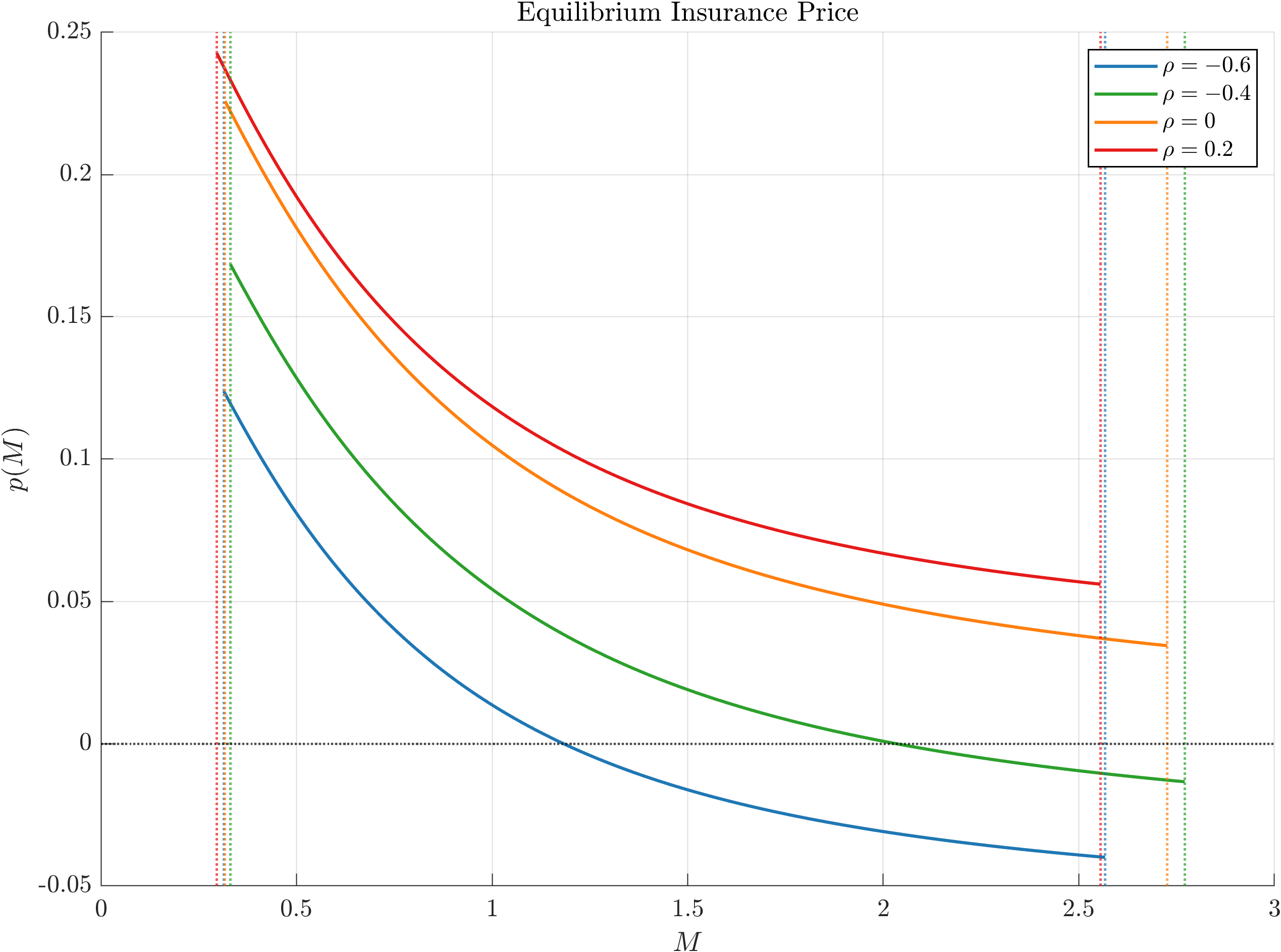}
    \includegraphics[width=0.45\linewidth]{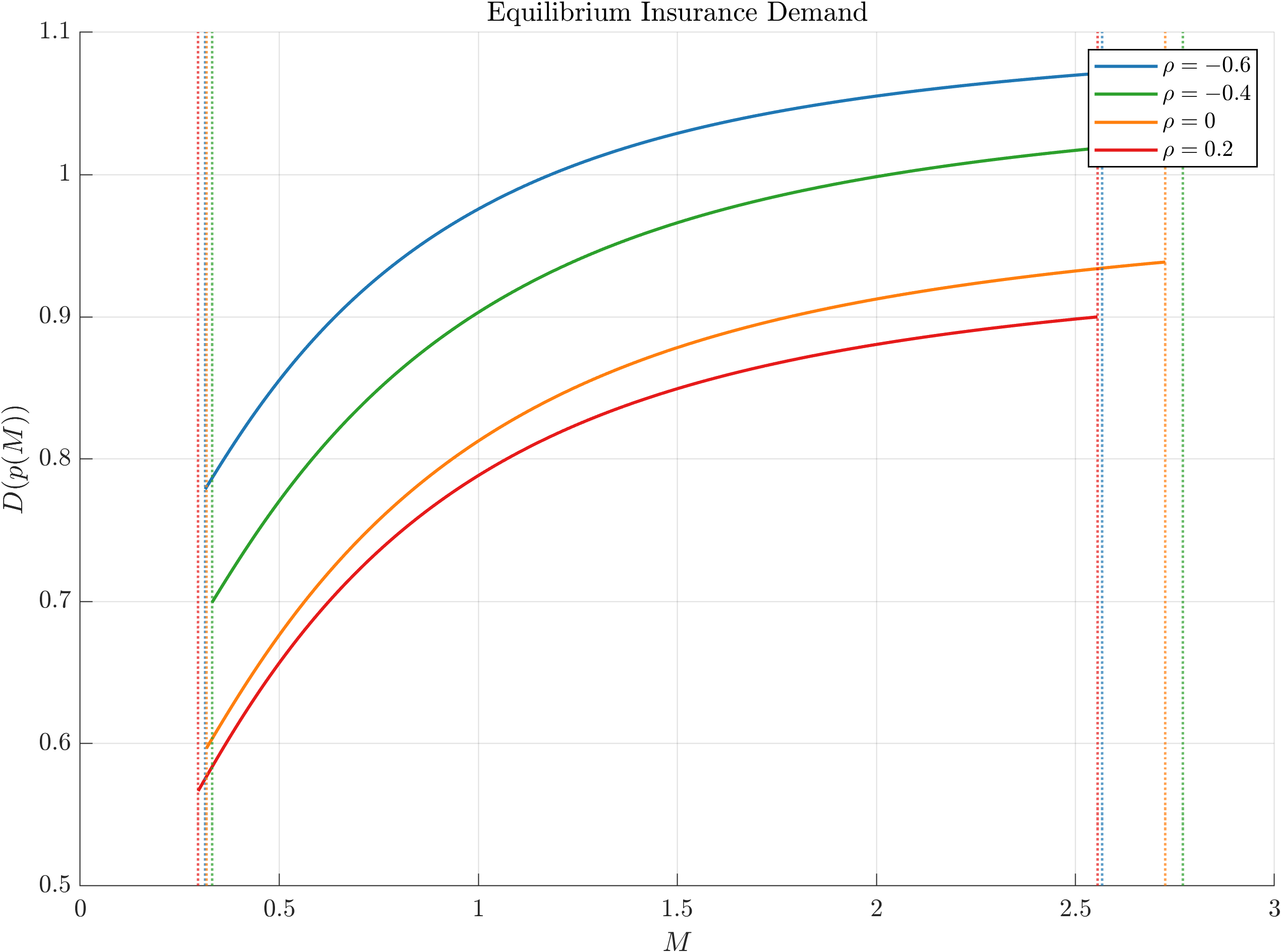}
    \includegraphics[width=0.45\linewidth]{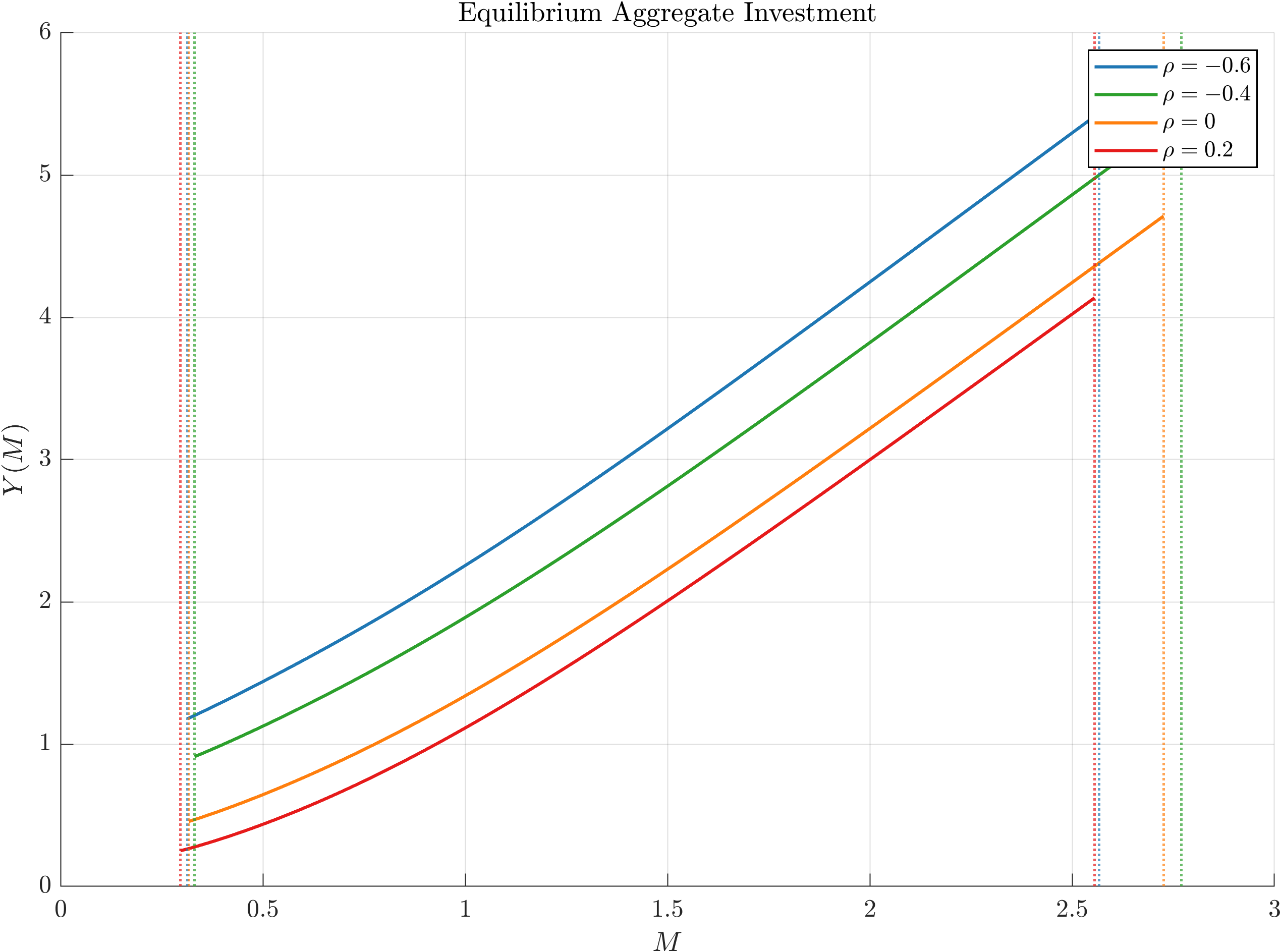}
    \caption{Equilibrium Outcomes for Different Correlation Coefficients $\rho\in\{-0.60,-0.40,0,0.20\}$.}
    \label{Figure Different Correlation}
\end{figure}

Fixing all other parameters, we vary the correlation coefficient to examine its impact on equilibrium outcomes, as shown in Figure~\ref{Figure Different Correlation}. The reported admissible cases use $\rho\in \{-0.60,-0.40,0,0.20\}$. Consistent with Proposition~\ref{Proposition sign of loading}, nonnegative correlation does not generate negative loading. By contrast, sufficiently negative correlation can generate a negative-loading region when aggregate capacity is high.

Overall, higher values of $\rho$ shift the equilibrium price curves upward and reduce aggregate financial investment. These patterns are consistent with the investment hedging channel: negative dependence between underwriting surplus and financial returns makes financial investment more valuable as a hedge, which is reflected in lower equilibrium insurance prices and higher investment exposure. Besides, the market-to-book ratio preserves its decreasing shape across all cases, suggesting that the barrier structure remains stable even though prices, demand, investment, and endogenous boundaries adjust with the correlation coefficient.  

\subsection{Impact of Robustness Degree}

\begin{figure}[t]
    \centering
    \includegraphics[width=0.45\linewidth]{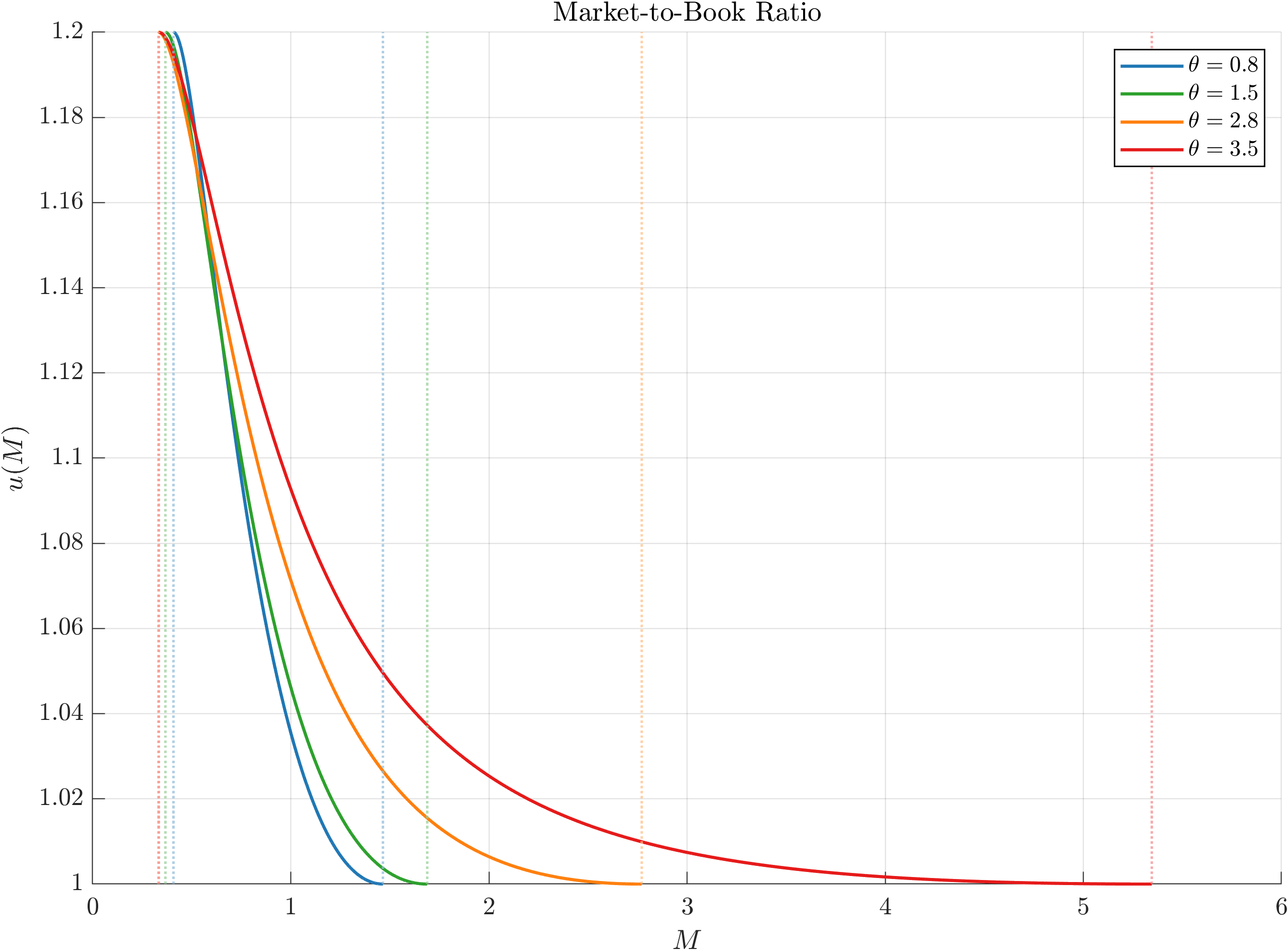}
    \includegraphics[width=0.45\linewidth]{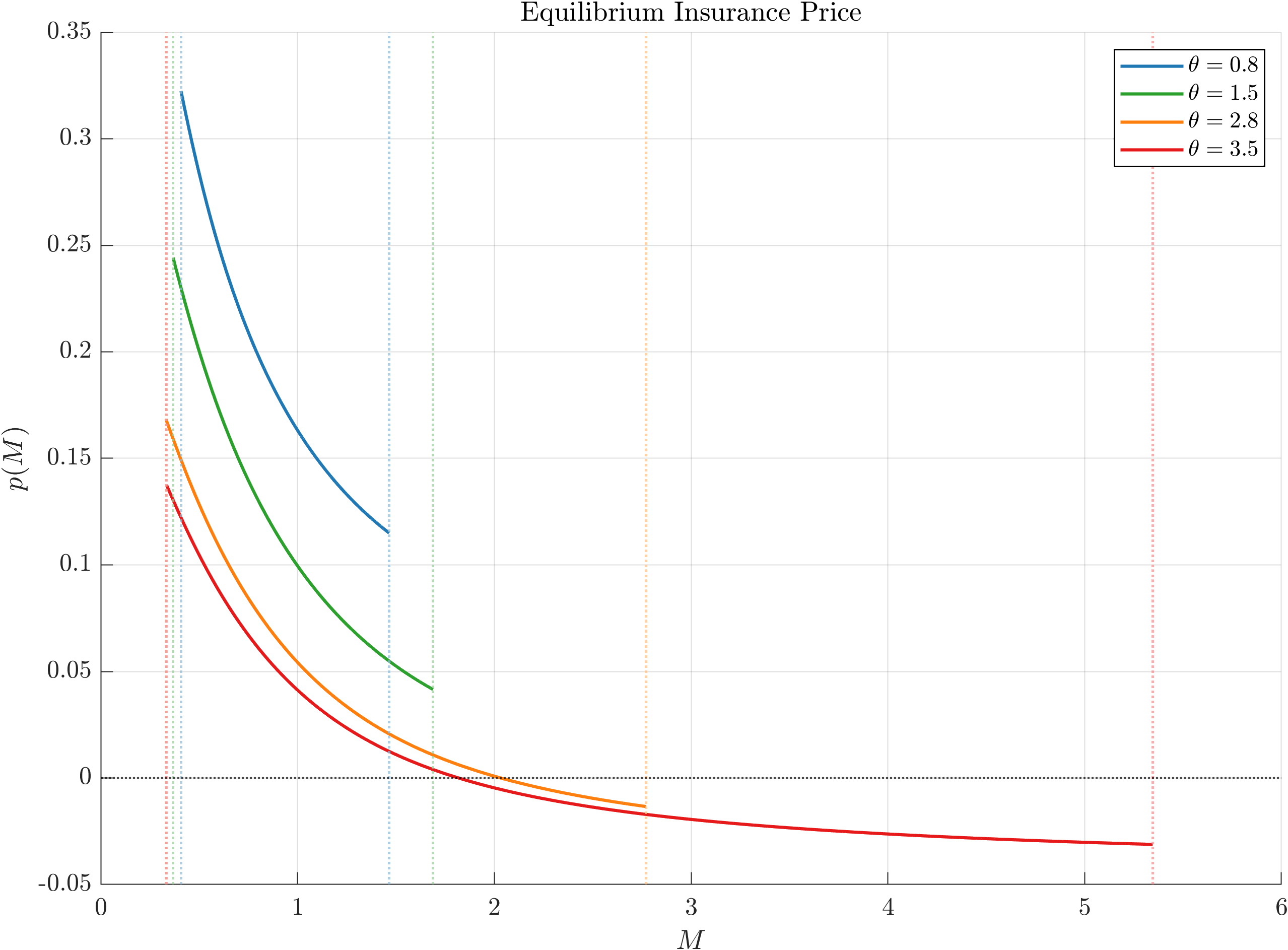}
    \includegraphics[width=0.45\linewidth]{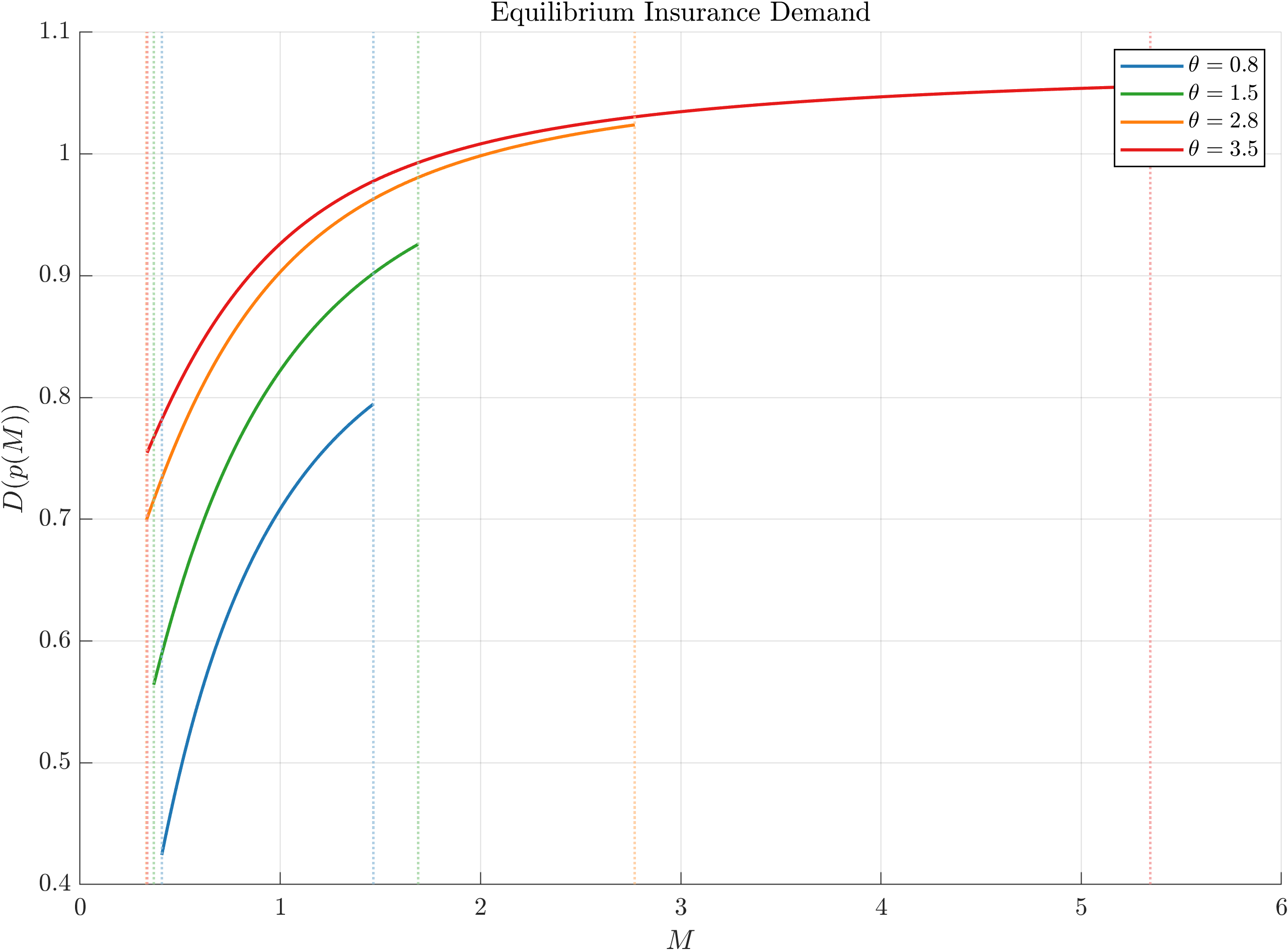}
    \includegraphics[width=0.45\linewidth]{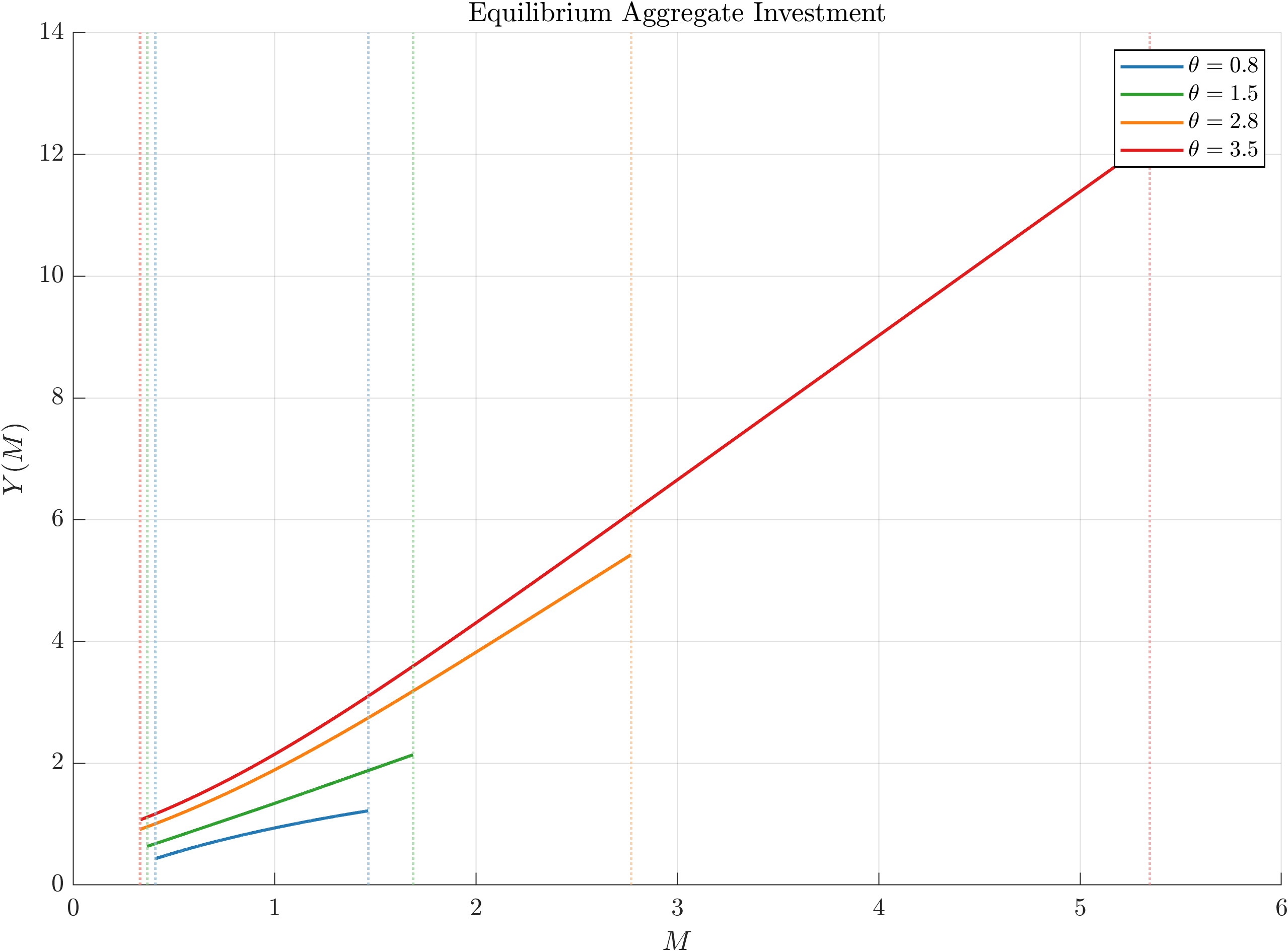}
    \caption{Equilibrium Outcomes for Stable Robustness Degrees $\theta\in\{0.8,1.5,2.8,3.5\}$.}
    \label{Figure Different Robustness Degree} 
\end{figure}

We next examine how the degree of concern for model uncertainty affects equilibrium outcomes within the admissible parameter region. Theoretically, a larger $\theta$ corresponds to lower ambiguity aversion, implying that insurers place greater confidence in the reference probability model. Figure~\ref{Figure Different Robustness Degree} reports the results for values of $\theta\in \{0.8,1.5,2.8,3.5\}$. For a given level of aggregate capacity, the equilibrium insurance price decreases as $\theta$ increases from a moderately low to a moderately high level. This pattern is consistent with the standard interpretation of robust control: when insurers are more concerned about model misspecification, they require higher compensation for bearing underwriting and financial risks. 

The robustness parameter also affects the endogenous boundaries. As $\theta$ increases, the payout boundary expands substantially, indicating that dividend payments are delayed. This behavior differs from the model without external financial investment in \citet{pang2026robust}, where the payout boundary eventually converges to its no-ambiguity counterpart as $\theta\to\infty$. In the present model, the investment channel prevents such a smooth convergence. As ambiguity concerns weaken, the system moves toward the no-ambiguity investment benchmark, in which the investment rule becomes ill-defined near a payout boundary satisfying $R(\overline M)=0$. 

Therefore, the robust verification system should not be read as a global existence result. The ODE system in Proposition~\ref{Proposition Robust Equilibrium} provides a set of necessary conditions for a stationary Markovian barrier equilibrium under Assumptions~\ref{Assumption market-to-book ratio} and~\ref{Assumption Robust}, but we do not prove existence or uniqueness of its solution for all parameter values. In the numerical implementation, the benchmark calibration produces an admissible verification solution, and the targeted theta audit suggests that this solution remains stable only over an intermediate range of robustness parameters. In particular, under the benchmark parameter setting, the case $\theta=3.8$ is close to the boundary of numerical admissibility, while $\theta=4.0$ and larger values in the targeted grid fail to produce admissible verification solutions. 

\begin{Remark}
    An important theoretical contribution of this paper is to demonstrate the existence of an equilibrium within a dynamic liquidity management framework that incorporates outside financial investment opportunities. While we acknowledge the contribution of \citet{luciano2022fluctuations}, which extends the benchmark model of \citet{henriet2016dynamics} by introducing a financial market to explain certain asset pricing phenomena, we point out that the equilibrium they define (Theorem 1 in their paper) does not actually exist, at least not in a barrier-type form. More specifically, there cannot exist a finite payout boundary satisfying the condition $R(\overline{M}) = 0$. Otherwise, according to their formulation, the aggregate investment amount $Y^{\ast}(M) = \frac{\mu - r}{R^{\ast}(M)\sigma^2}$ would be ill-defined at $\overline{M}$, and consequently, the ODE satisfied by the market-to-book ratio function would have no solution consistent with the boundary conditions. 

    In our setting, by additionally incorporating a correlation term $\rho$, if we disregard model uncertainty, we obtain a similar expression $Y^{\ast}(M) = \frac{1}{1 - \rho^2} \frac{q - \rho p^{\ast}(M)}{R^{\ast}(M) \sigma}$, which again cannot be well-defined under the condition $R(\overline{M}) = 0$, implying that the equilibrium does not exist in this case either. By contrast, once model uncertainty is introduced, our numerical results indicate that under reasonable parameter configurations, the non-existence problem can be mitigated. 
    
    The limit $\theta\to\infty$ clarifies the connection with the no-ambiguity benchmark. On any fixed compact state interval $[\underline M,\overline M]$ with $\underline M>0$ and bounded $u(M)$ and $R(M)$, 
    \begin{equation}
        A_{\theta}(M)=R(M)+\frac{u(M)}{\theta}\left(\frac{1}{M}-R(M)\right)
        \xrightarrow[]{\theta\rightarrow\infty}R(M). \notag
    \end{equation}
    Consequently, the pricing and investment subsystem converges pointwise to the benchmark form
    \begin{align}
        p^{\ast}(M)&\xrightarrow[]{\theta\rightarrow\infty}
        \frac{R(M)(1-\rho^2)\eta+\rho q}
        {1+ \frac{1}{\alpha} R(M)(1-\rho^2)}, \notag\\
        Y^{\ast}(M)&\xrightarrow[]{\theta\rightarrow\infty}
        \frac{1}{1-\rho^2}\frac{q-\rho p^{\ast}(M)}{R(M)\sigma}. \notag
    \end{align}
    The limiting investment rule is well defined only at states where $R(M)\neq0$. In the robust verification system, the finite-boundary conditions imply $R(\underline M)=R(\overline M)=0$, so the no-ambiguity limit cannot be evaluated at either boundary. This feature should be distinguished from the no-ambiguity benchmark, where the lower boundary is typically given by $\underline M=0$ rather than by $R(\underline M)=0$. Hence, the relevant obstruction in the finite-barrier investment model is the payout boundary: if $\overline M$ is finite and $R(\overline M)=0$, the limiting aggregate investment rule becomes singular. 
    
    The economic intuition is straightforward. At a finite payout boundary, the marginal value of additional internal funds coincides with the value of immediate dividend payments, so the implied market risk aversion $R(M)$ vanishes. In the absence of model uncertainty or other investment frictions, insurers therefore behave locally as if they were risk neutral with respect to the financial asset. Since the risky asset has a positive Sharpe ratio, they would prefer to allocate additional liquid reserves to financial investment rather than distribute them as dividends. This is why the aggregate investment rule becomes unbounded in the no-ambiguity limit. Introducing model uncertainty adds an informational friction: large financial positions also expose insurers to model misspecification concerns, which restrains investment demand and helps restore a finite-barrier verification solution over an intermediate range of robustness parameters. 
\end{Remark}

\begin{table}[htbp]
\centering
\setlength{\tabcolsep}{8pt}
\renewcommand{\arraystretch}{1.4}
\small
\begin{tabular}{cccc}
\hline
Parameter       & \multicolumn{1}{c}{External Financing Boundary, $\underline{M}$} & Payout Boundary, $\overline{M}$ & Range of Capacity, $\Delta{M}$ \\ \hline
$\rho = -0.6$    & 0.3139                                                            & 2.5670                          & 2.2531                         \\
$\rho = -0.4$    & 0.3312                                                            & 2.7705                          & 2.4394                         \\
$\rho = 0$       & 0.3171                                                            & 2.7259                          & 2.4087                         \\
$\rho = 0.2$     & 0.2962                                                            & 2.5554                          & 2.2592                         \\ \hline
$\theta = 0.8$   & 0.4079                                                            & 1.4655                          & 1.0576                         \\
$\theta = 2.8$   & 0.3312                                                            & 2.7705                          & 2.4394                         \\
$\theta = 3.5$   & 0.3345                                                            & 5.3456                          & 5.0111                         \\
$\theta = 3.8$   & 0.3448                                                            & 10.7462                         & 10.4014                        \\ \hline
$\gamma = 0.02$  & 0.6698                                                            & 1.9380                          & 1.2682                         \\
$\gamma = 0.1$   & 0.4312                                                            & 2.4729                          & 2.0417                         \\
$\gamma = 0.2$   & 0.3312                                                            & 2.7705                          & 2.4394                         \\
$\gamma = 0.3$   & 0.2775                                                            & 2.9519                          & 2.6745                         \\ \hline
\end{tabular}
\caption{Numerically Solved Boundary Values.}
\label{Table Boundary Values}
\end{table}

\subsection{Impact of Financing Cost}

\begin{figure}[t]
    \centering
    \includegraphics[width=0.45\linewidth]{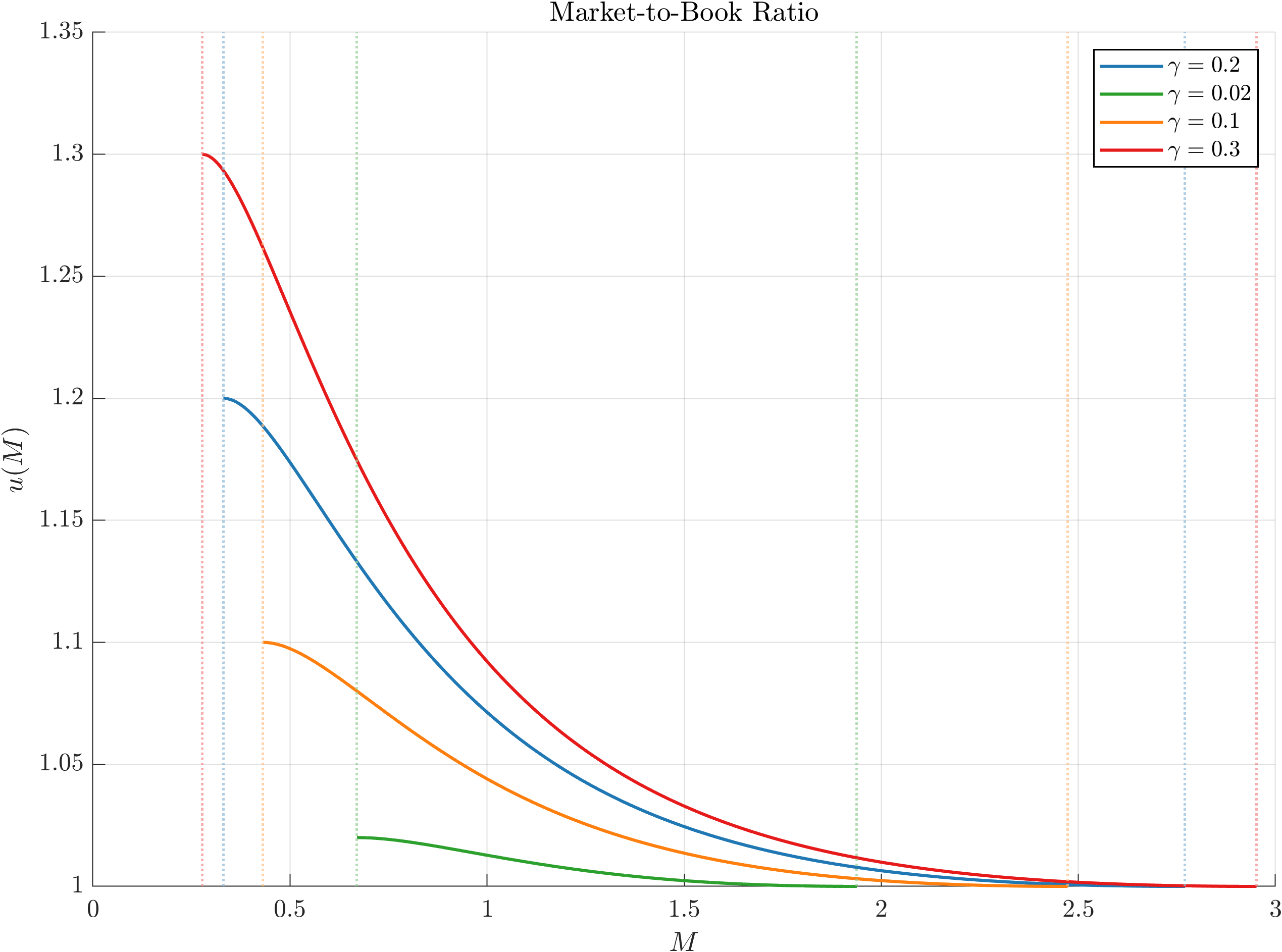}
    \includegraphics[width=0.45\linewidth]{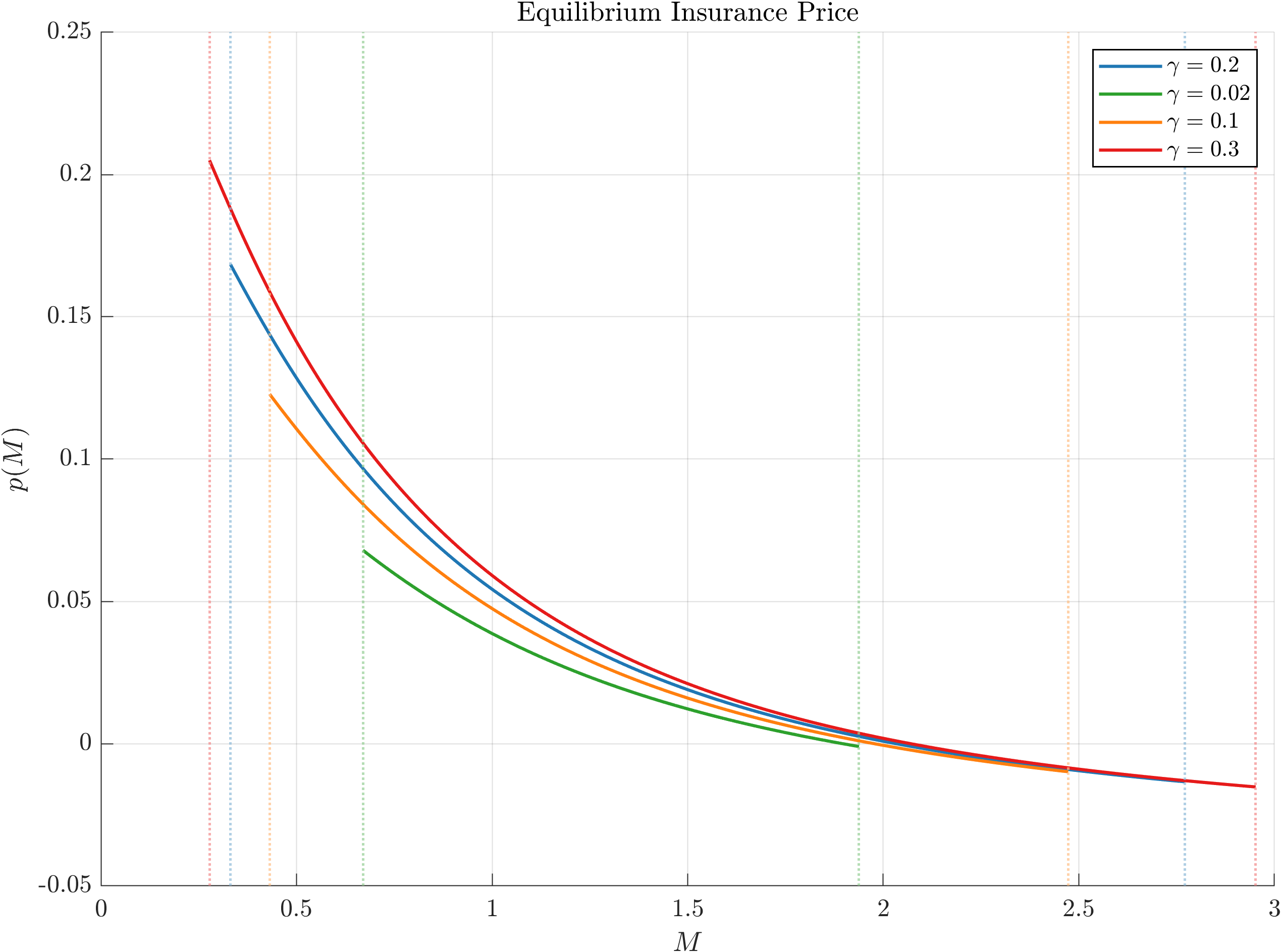}
    \includegraphics[width=0.45\linewidth]{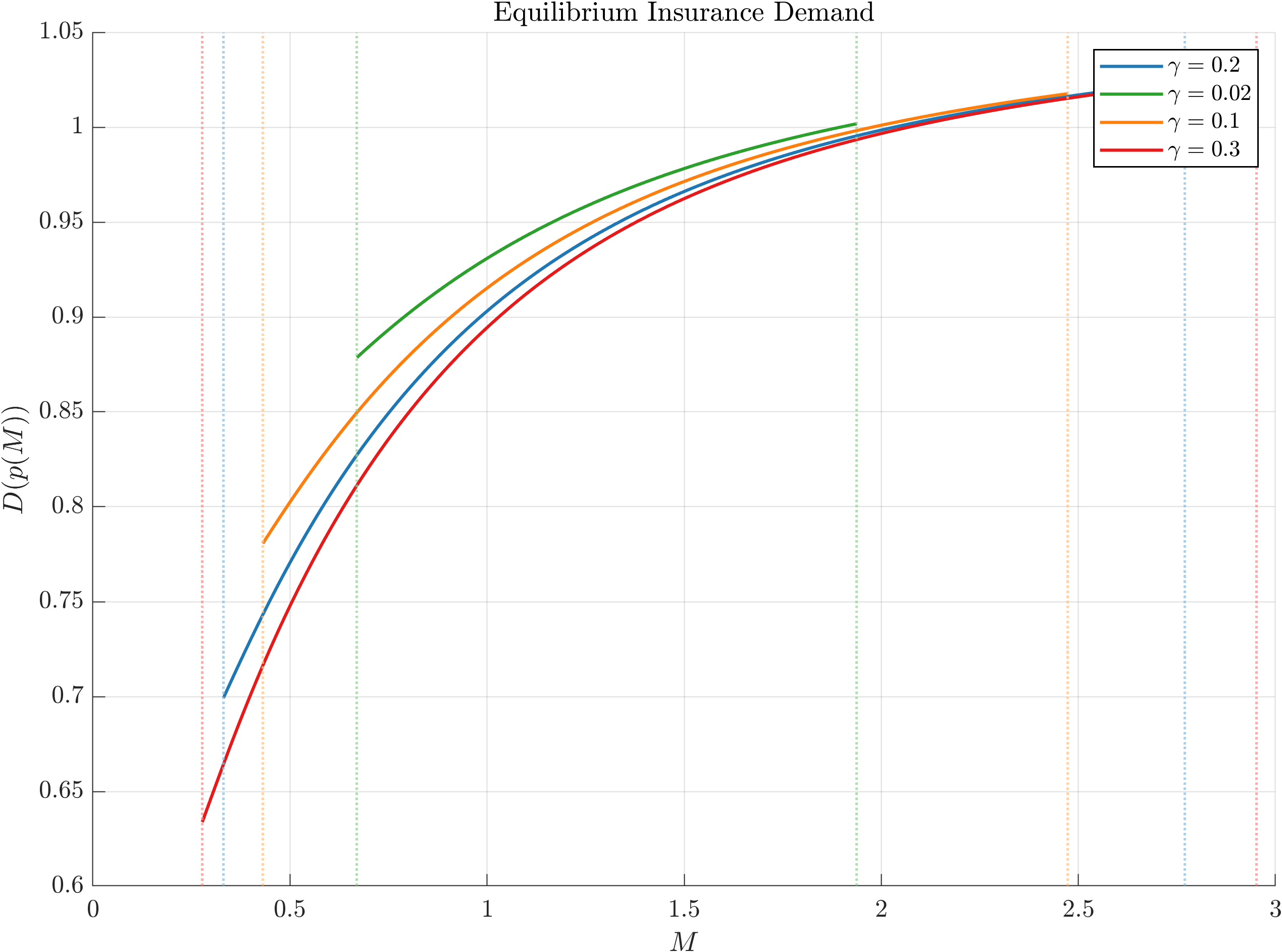}
    \includegraphics[width=0.45\linewidth]{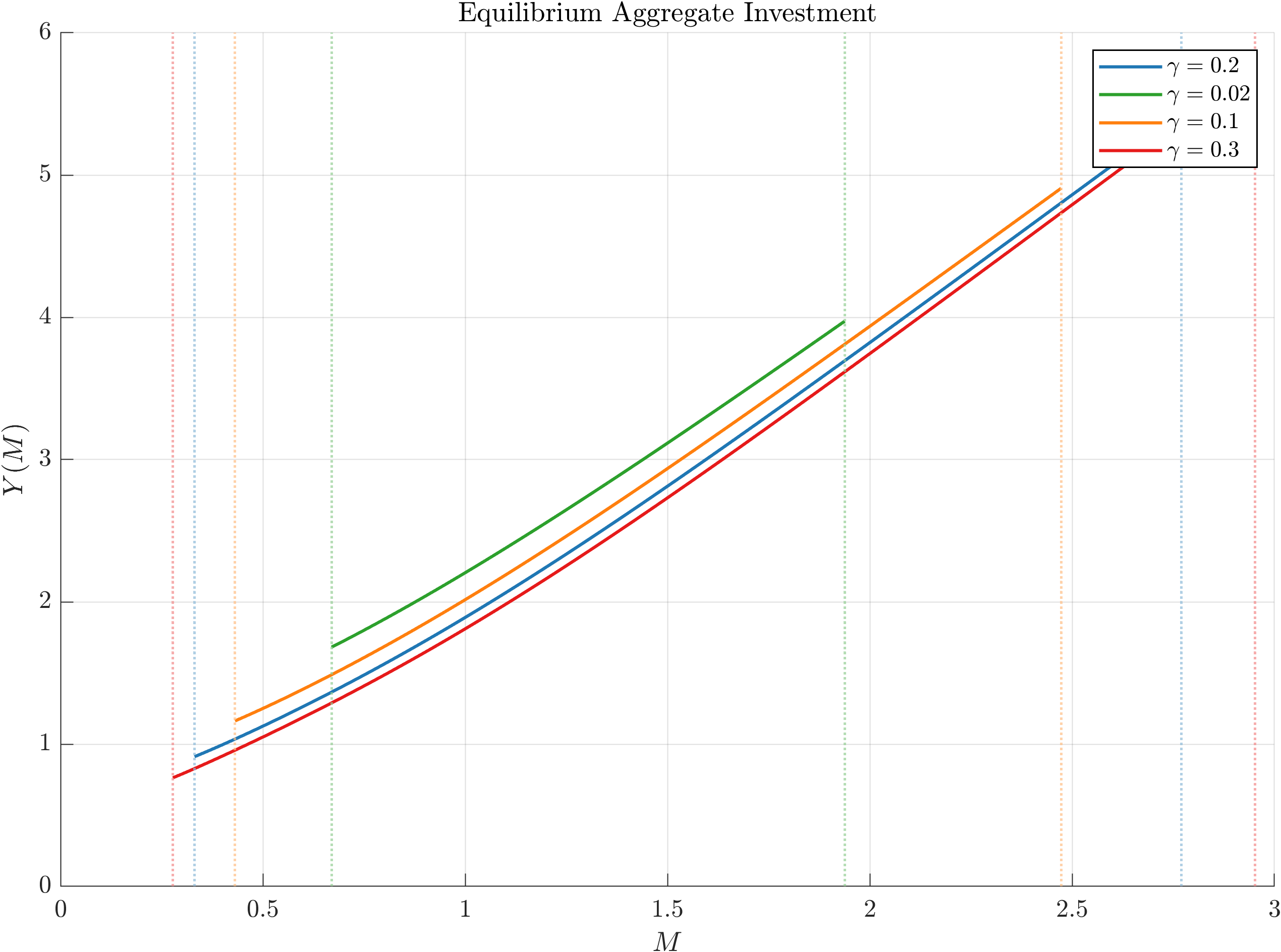}
    \caption{Equilibrium Outcomes for Different Financing Cost $\gamma$.}
    \label{Figure Different Financing Cost}
\end{figure}

We further examine how external financing costs affect equilibrium outcomes. Figure~\ref{Figure Different Financing Cost} shows that a higher $\gamma$ lowers the recapitalization boundary and raises the payout boundary, thereby widening the endogenous capacity region. Intuitively, when external equity becomes more costly, insurers rely more on internal liquidity: they tolerate lower reserves before raising new capital and postpone dividend payouts in high-capacity states. The market-to-book ratio also shifts upward with $\gamma$, reflecting the higher value of internal capital when external financing is expensive.

The pricing and allocation effects are consistent with this interpretation. On the overlapping part of the state space, the equilibrium insurance price increases with $\gamma$, while insurance demand decreases accordingly. This monotonic pricing pattern follows from the fact that underwriting uses scarce internal capital, whose opportunity cost is higher when recapitalization is more expensive. Aggregate risky investment also tends to be lower when financing costs are higher, as insurers become more cautious in allocating internal capital to financial risk-taking. Overall, financing frictions raise the value of liquidity, increase equilibrium loadings, reduce insurance demand, and expand the range over which insurers manage capital internally. 

\section{Insurers’ Long-Run Behavior Pattern} \label{Section Long-Run Behavior}

In this section, we simulate the insurance market equilibrium in a dynamic setting to gain deeper insights into the long-run behavior of insurers. In particular, we analyze the cyclical patterns of underwriting activity and the ergodic properties of the insurance market over time. 

\subsection{Underwriting Dynamics}

\begin{figure}[htbp]
    \centering
    \includegraphics[width=0.45\linewidth]{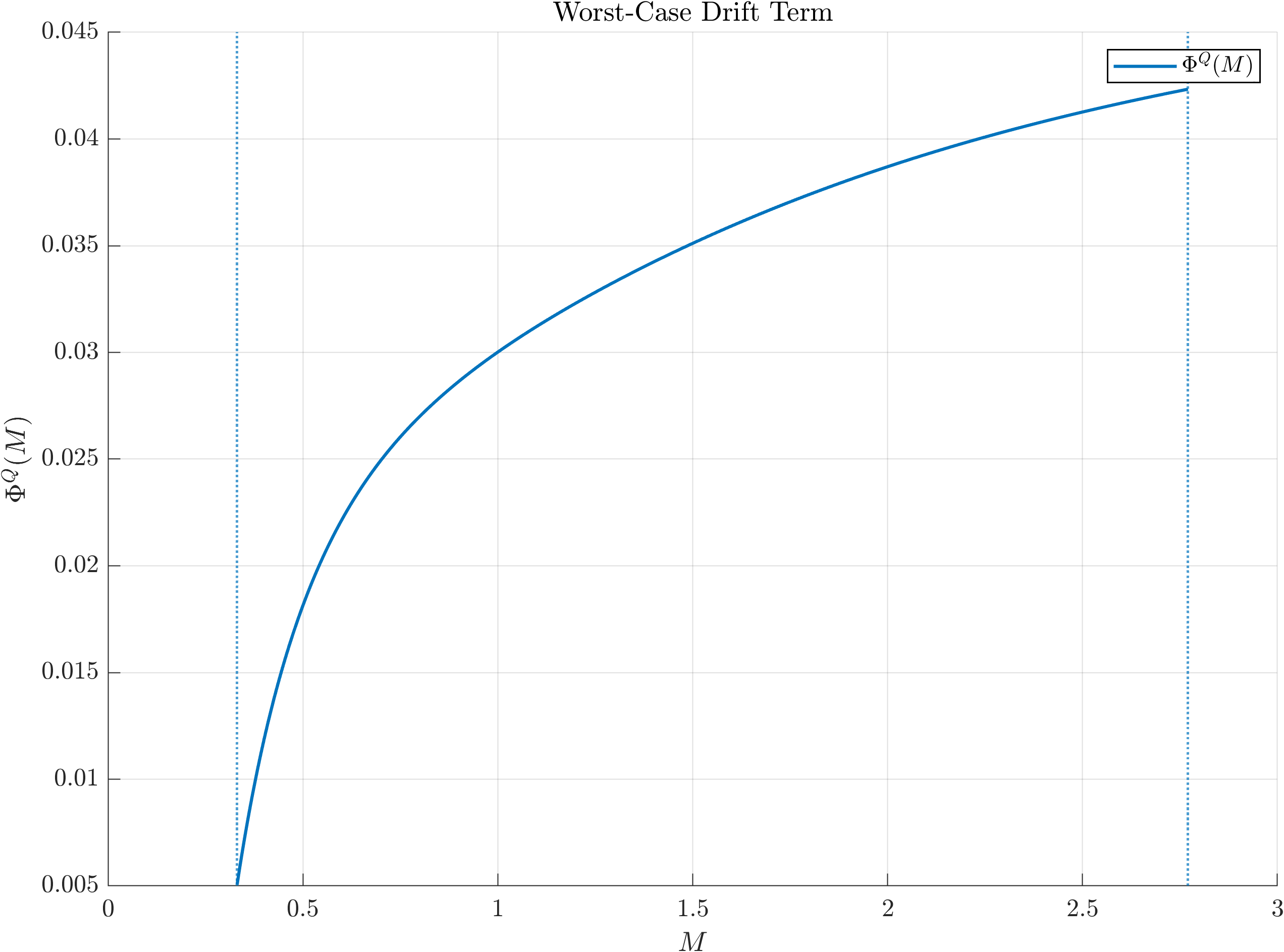}
    \includegraphics[width=0.45\linewidth]{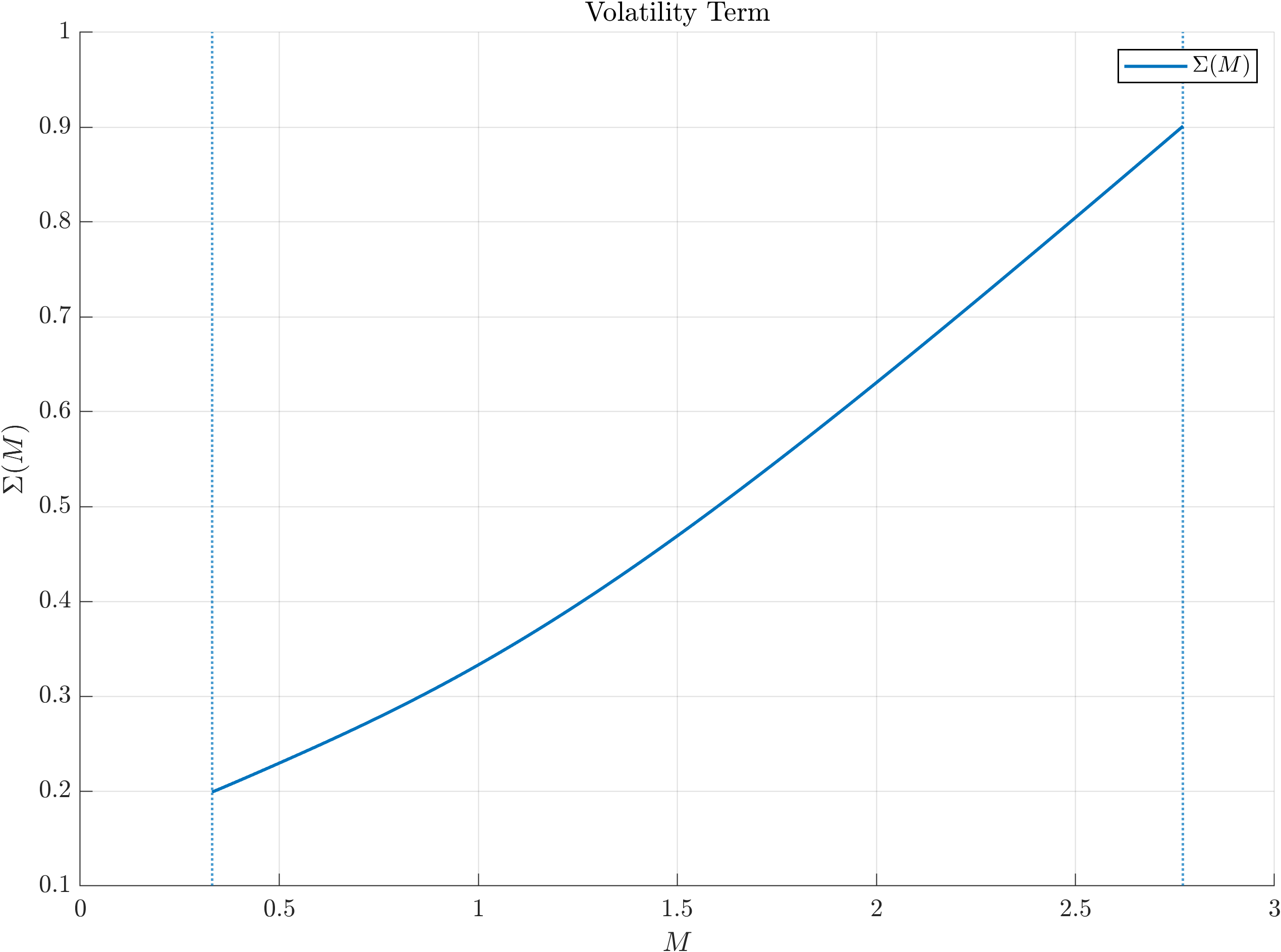}
    \caption{Drift and Volatility Terms of Dynamic Liquid Reserves Process.} 
    \label{Figure Dynamics Drift and Volatility}
\end{figure}

Given the optimally chosen underwriting and pricing strategies, the aggregate capacity of the insurance sector evolves on $[\underline{M}, \overline{M}]$ according to: 
\begin{align}
    \mathrm{d}M_{t} = & \left[ D\!\left(p^{\ast}(M_{t})\right) \eta \Big(p^{\ast}(M_{t}) + {h^I}^{\ast}(M_{t}) \Big) + Y^{\ast}\!\left(M_{t}\right) \sigma \Big(q + {h^S}^{\ast}(M_{t}) \Big) + M_t r \right] \,\mathrm{d}t \notag \\
    & + D\!\left(p^{\ast}(M_{t})\right) \eta \,\mathrm{d}W_t^{I, {h}^{\ast}} + Y^{\ast}\!\left(M_{t}\right) \sigma \,\mathrm{d}W_t^{S, {h}^{\ast}} \notag \\
    = & \underbrace{ \bigg\{ M_t r + R^{\ast}(M_{t}) \bigg[ \Big(D(p^{\ast}(M_{t})) \eta \Big)^2 + 2 \rho D(p^{\ast}(M_{t})) \eta Y^{\ast}(M_t) \sigma + \Big(Y^{\ast}(M_t) \sigma \Big)^2  \bigg] \bigg\}}_{\varPhi(M)} \,\mathrm{d}t 
    \notag \\
    & + \underbrace{ \bigg[ \Big(D(p^{\ast}(M_{t})) \eta \Big)^2 + 2 \rho D(p^{\ast}(M_{t})) \eta Y^{\ast}(M_t) \sigma + \Big(Y^{\ast}(M_t) \sigma \Big)^2  \bigg]^{\frac{1}{2}}}_{\varSigma(M)} \,\mathrm{d}W_t^{h^{\ast}},    
    \label{Equilibrium M Dynamics}
\end{align}
where $W_t^{I, {h}^{\ast}}$ and $W_t^{S, {h}^{\ast}}$ are Brownian motions under the worst-case measure $\mathbb{Q}^{h^{\ast}}$, and their linear combination $W_t^{h^{\ast}}$ is also a Brownian motion. Figure~\ref{Figure Dynamics Drift and Volatility} plots the drift and volatility terms in~\eqref{Equilibrium M Dynamics}. While $\varPhi(M)$ remains strictly positive, its magnitude is substantially smaller than $\varSigma(M)$. This implies that the capacity dynamics are primarily driven by the diffusion component. 

\begin{figure}[t]
    \centering
    \includegraphics[width=0.9\linewidth]{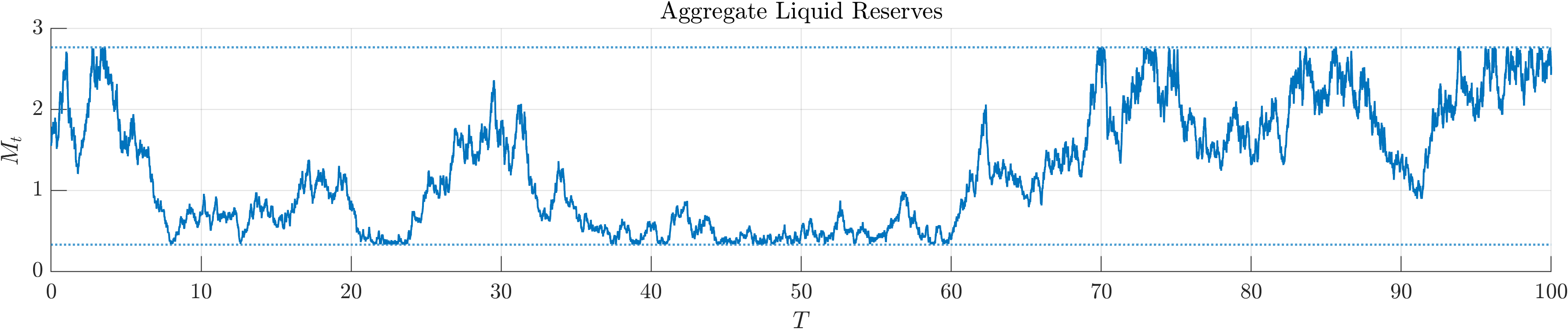}
    \includegraphics[width=0.9\linewidth]{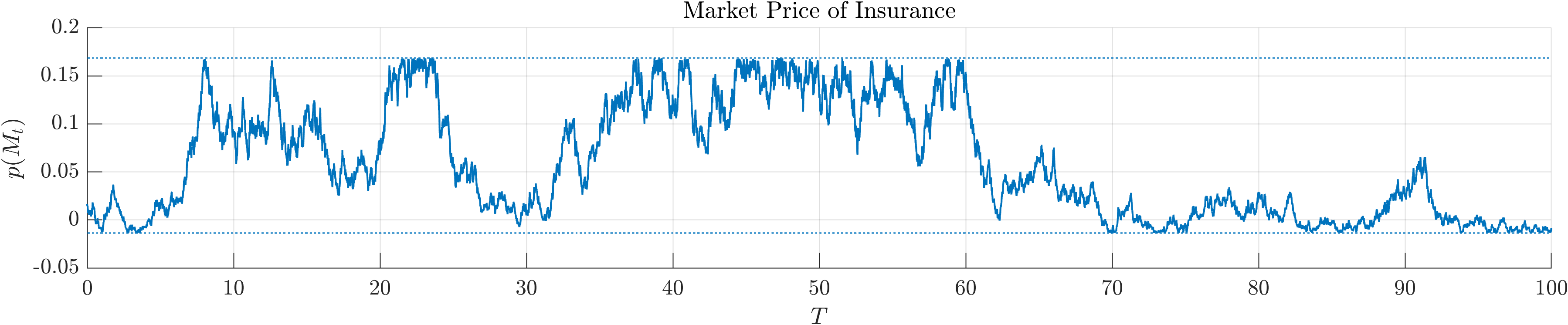}
    \includegraphics[width=0.9\linewidth]{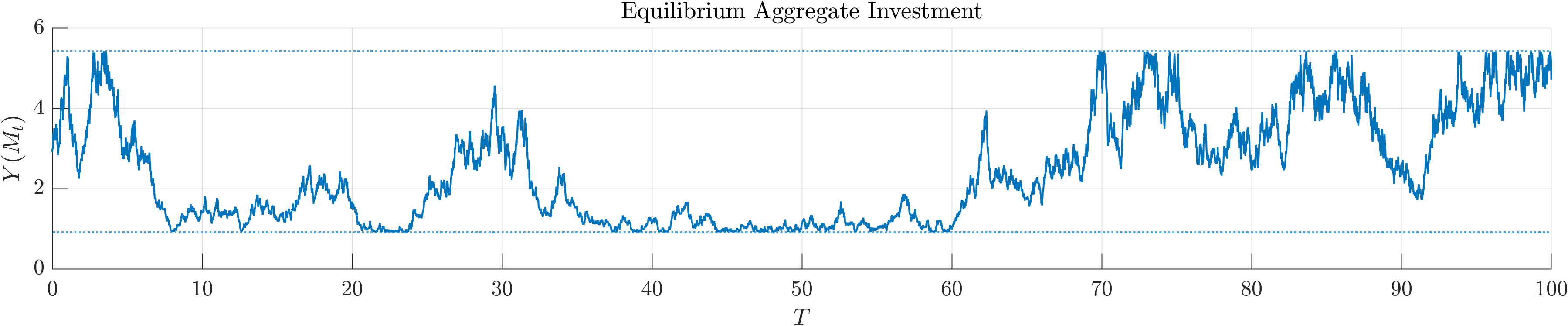}
    \caption{Dynamic Equilibrium Outcomes.} 
    \label{Figure Dynamics Comparison}
\end{figure}

Figure~\ref{Figure Dynamics Comparison} presents the simulated dynamic equilibrium outcomes of the insurance market, including the evolution of aggregate reserves, equilibrium insurance prices, and aggregate investment. The insurance price adjusts endogenously with the level of aggregate capacity, generating cyclical fluctuations that resemble the well-documented underwriting cycles observed in insurance markets \citep{harrington2013insurance}. Insurers’ holdings of risky assets also display cyclical variation, consistent with the ``flight-to-quality'' phenomenon, whereby insurers reallocate their portfolios toward safer assets and away from riskier ones when they become financially constrained.  

\subsection{Duration of Underwriting Cycles}

Following \citet{henriet2016dynamics}, we further compute the expected duration of underwriting cycles, which are characterized by alternating phases of: (1) a soft market, during which insurers’ aggregate capacity expands from $\underline{M}$ to $\overline{M}$, leading to declining equilibrium prices; and (2) a hard market, during which aggregate capacity contracts from $\overline{M}$ back to $\underline{M}$, accompanied by rising prices. Formally, let $T_s(M)$ denote the expected time for the reserve process $M_t$ to reach the upper boundary $\overline{M}$ from any state $M \leq \overline{M}$. Then $T_s(\underline{M})$ represents the expected duration of the soft-market phase. Similarly, let $T_h(M)$ denote the expected time for $M_t$ to return from any state $M \geq \underline{M}$ to the lower boundary $\underline{M}$. Then $T_h(\overline{M})$ represents the expected duration of the hard-market phase. Finally, the total expected duration of an insurance cycle is defined as $T_c \triangleq T_s(\underline{M}) +  T_h(\overline{M})$.  

\begin{Proposition}
    Let the aggregate capacity $M_{t}$ evolve on $[\underline M, \overline M]$ with drift $\varPhi(\cdot)$ and volatility $\varSigma(\cdot)$. Then, $T_{s}(\cdot)$ and $T_{h}(\cdot)$ should solve the following ODEs:  
    \begin{align}
        -1 & = \varPhi(M) T^{\prime}_{s}(M) + \frac{1}{2} \varSigma(M)^2 T^{\prime \prime}_{s}(M), \quad \text{with} \quad T^{\prime}_{s}(\underline{M}) = 0, \ T_{s}(\overline{M}) = 0, \notag \\
        -1 & = \varPhi(M) T^{\prime}_{h}(M) + \frac{1}{2} \varSigma(M)^2 T^{\prime \prime}_{h}(M), \quad \text{with} \quad T_{h}(\underline{M}) = 0, \ T^{\prime}_{h}(\overline{M}) = 0. \notag
    \end{align}
    The Neumann conditions impose instantaneous reflection at the non-target boundary, while the Dirichlet conditions set the hitting time to zero upon arrival at the target boundary.
\end{Proposition} 

\begin{figure}[t]
    \centering
    \includegraphics[width=0.45\linewidth]{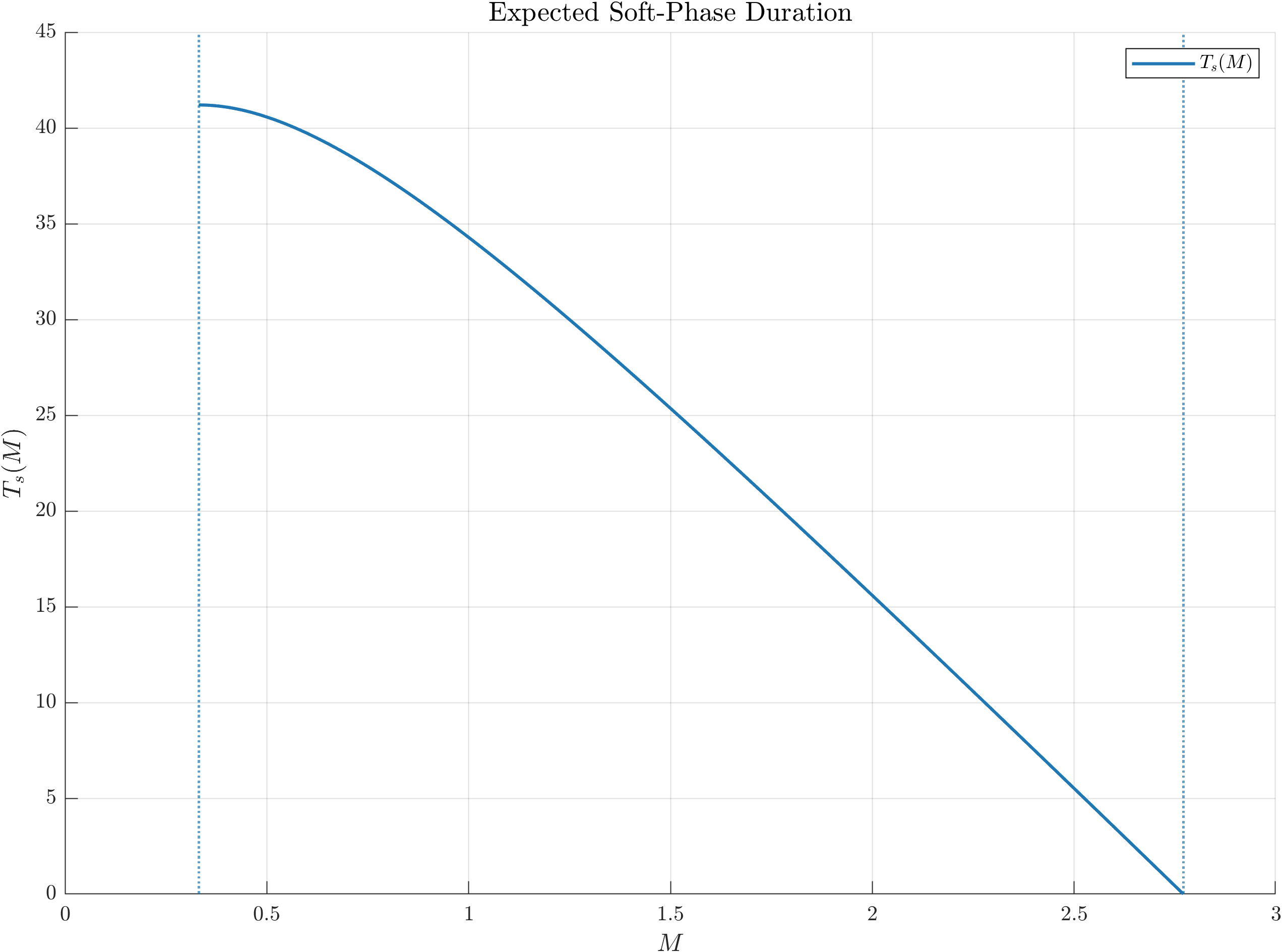}
    \includegraphics[width=0.45\linewidth]{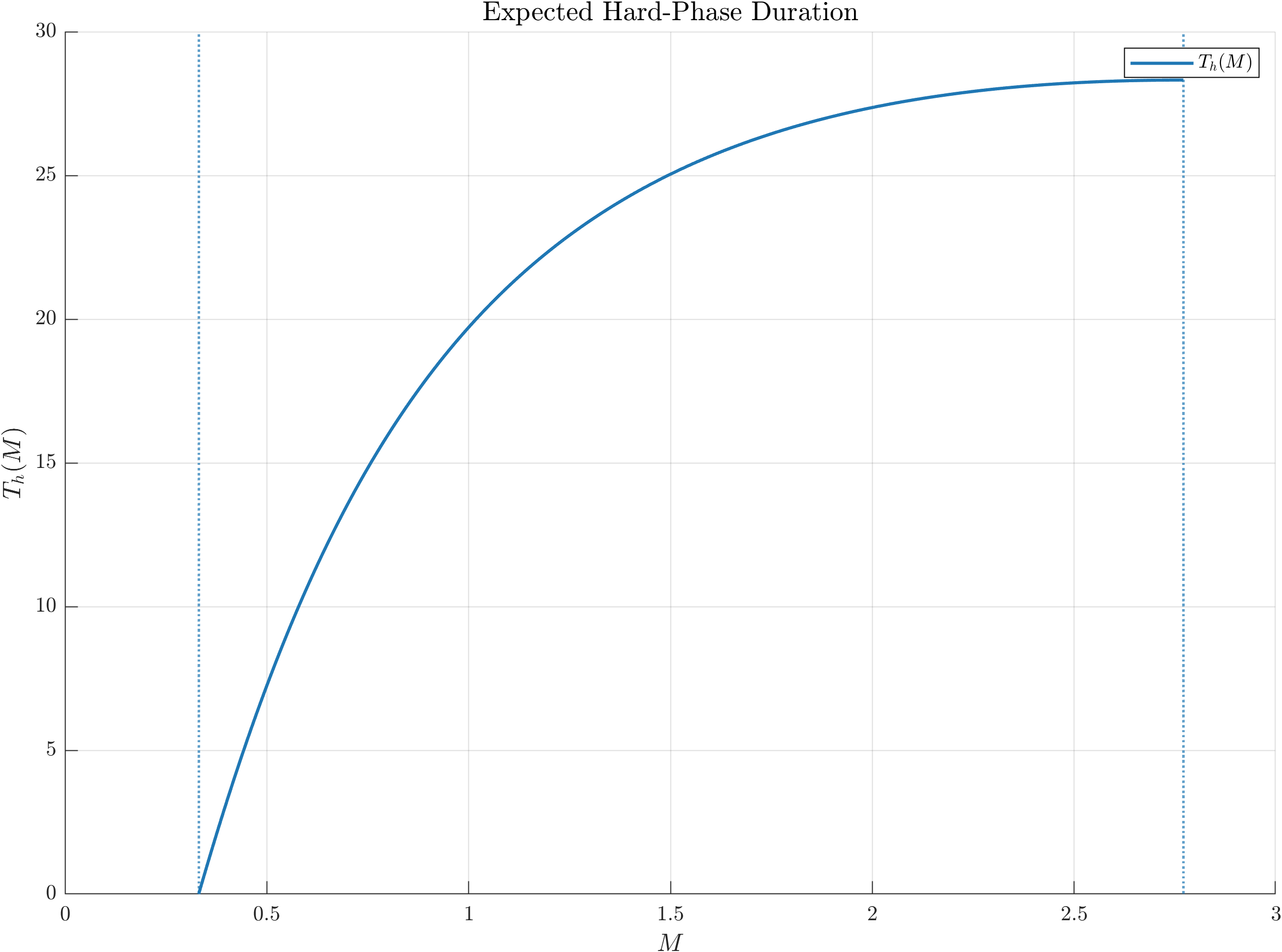}
    \caption{Expected Soft and Hard Phase Durations.} 
    \label{Figure Duration}
\end{figure} 

The result follows from the Feynman-Kac representation for diffusions with reflecting boundaries \citep[see, e.g.,][]{lions1984stochastic, karatzas2014brownian}. A detailed derivation in a closely related setting can be found in Section 4.2 of \citet{henriet2016dynamics}. Figure~\ref{Figure Duration} presents the numerical solutions for $T_{s}(\cdot)$ and $T_{h}(\cdot)$. It is estimated that $T_{s}(\underline{M}) = 41.20$ and $T_{h}(\overline{M}) = 28.32$, yielding a total expected cycle duration of $T_{c} = 69.52$. Consistent with empirical evidence \citep{henriet2016dynamics}, the soft market phase lasts longer than the hard market phase, reflecting insurers’ gradual capacity buildup and slower price adjustment during expansion periods. 

Notably, these durations are an order of magnitude longer than those obtained in models without financial investment. Under the same parameterization but without financial markets or model uncertainty, \citet{henriet2016dynamics}'s model estimates $T_{s}(\underline{M}) = 4.78$, $T_{h}(\overline{M}) = 4.84$, and $T_{c} = 9.62$. When robustness concern is introduced but financial investment is excluded, as in \citet{pang2026robust}, the corresponding durations increase to $T_{s}(\underline{M}) = 14.05$, $T_{h}(\overline{M}) = 11.92$, and $T_{c} = 25.97$. These comparisons highlight that the combination of model uncertainty and financial investment amplifies the persistence of underwriting cycles substantially, as insurers become more cautious in capacity adjustment and retain capital longer to buffer against uncertainty and market fluctuations. 

\subsection{Ergodic Property}

To further understand the long-run behavior of the insurance market, we examine the ergodic property of the equilibrium dynamics driven by fluctuations in aggregate capacity. Specifically, we study whether the capacity process $M_t$, bounded by the reflecting barriers $[\underline{M}, \overline{M}]$, converges to a stationary distribution over time. The existence of such a stationary measure implies that insurance cycles are statistically recurrent fluctuations around a long-run equilibrium state. 

\begin{Proposition}
The capacity process $\{M_{t}\}_{t \ge 0}$, evolving according to \eqref{Equilibrium M Dynamics}, admits a unique invariant probability measure $\pi$, and is ergodic in the sense that, for every bounded measurable function $f$,
\begin{equation}
    \lim_{T \rightarrow \infty} \frac{1}{T} \int_{0}^{T} f(M_{t}) \mathrm{d}t = \int_{\underline{M}}^{\overline{M}} f(M) \pi(\mathrm{d}M), \quad \text{a.s.}. \notag
\end{equation}
Moreover, $\pi$ admits a density $\tilde{\pi}(\cdot)$ that satisfies the stationary Fokker-Planck equation: 
\begin{equation}
    \tilde{\pi}(M) = \frac{\kappa}{\varSigma(M)^2 } \exp \left( 2 \int_{\underline{M}}^{M} \frac{\varPhi(z)}{\varSigma(z)^2} \mathrm{d}z  \right), \quad M \in [\underline{M}, \overline{M}], \notag
\end{equation}
where the normalizing constant is $\kappa^{-1}=\int_{\underline M}^{\overline M} \frac{1}{\varSigma(M)^2} \exp\!\Big( 2\int_{\underline M}^{M} \frac{\varPhi(z)}{\varSigma(z)^2} \mathrm dz \Big) \mathrm dM$. 
\end{Proposition} 

\begin{figure}[t]
    \centering
    \includegraphics[width=0.7\linewidth]{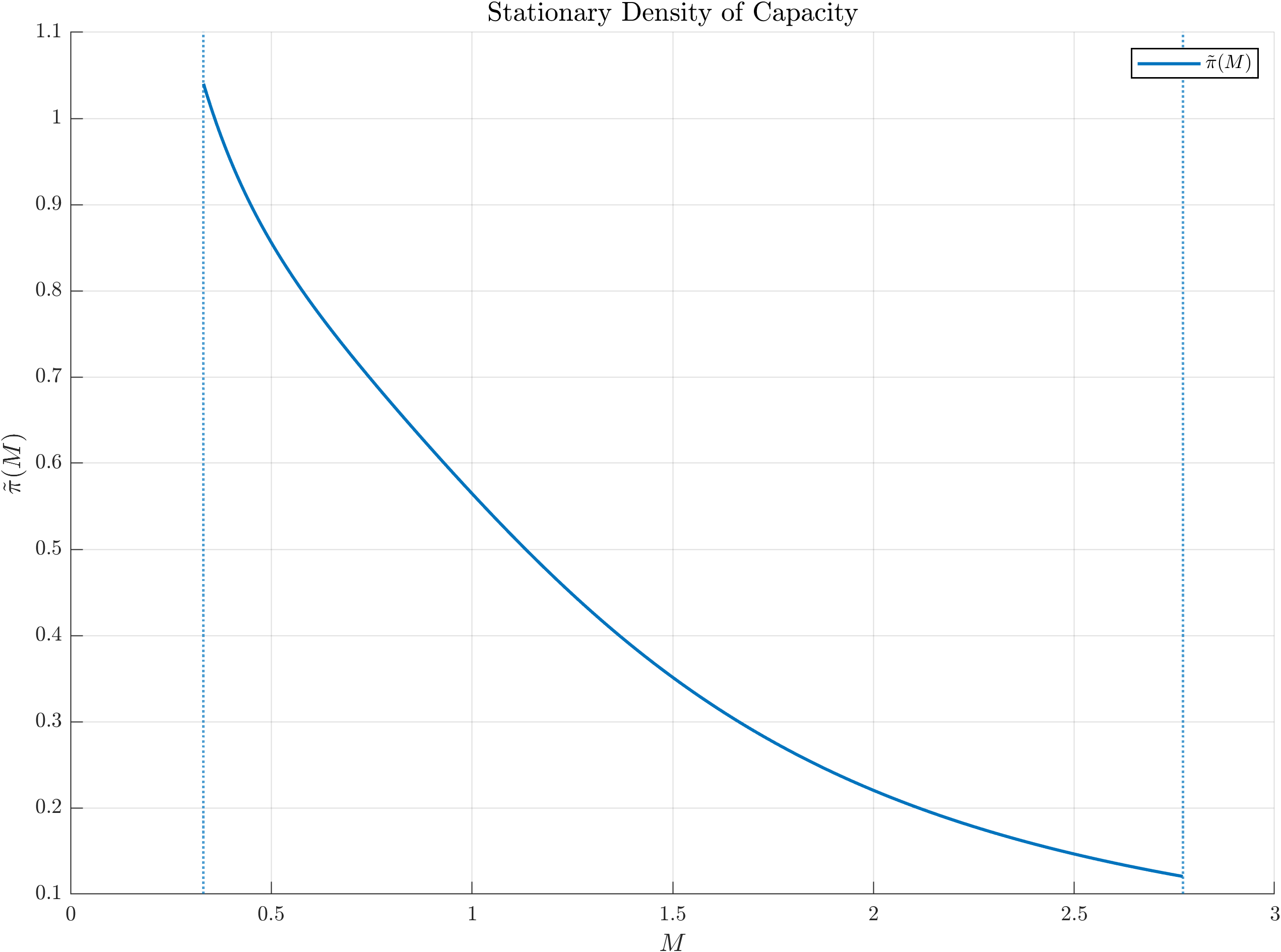}
    \caption{Stationary Density of Capacity.}
    \label{Figure Stationary Density}
\end{figure}

Figure~\ref{Figure Stationary Density} presents the numerically computed stationary density of the capacity process. The downward-sloping curve indicates that probability mass is concentrated at relatively low capacity levels. Due to the ergodic property, this implies that the insurance industry spends most of its time in capital-constrained states characterized by limited liquidity and conservative underwriting. This finding is consistent with the earlier evidence that soft phases tend to last longer than hard phases: since capacity typically builds up gradually but contracts abruptly, the stationary distribution becomes skewed toward low-capacity regimes. 

\section{Conclusion} \label{Section Conclusion}

This paper develops a dynamic equilibrium model of the insurance market that integrates insurers' underwriting, investment, recapitalization, and dividend decisions under the joint influence of model uncertainty and financial frictions. By embedding robust control into a barrier-type liquidity management framework, the analysis characterizes a stationary Markovian candidate equilibrium in which insurance pricing, financial investment, and liquidity management are jointly determined by aggregate capacity. Model uncertainty acts as an informational friction on insurers' risk-taking behavior and helps regularize the finite-barrier verification system under economically reasonable parameterizations, although it does not guarantee equilibrium existence for all parameter values. The resulting equilibrium generates liquidity-driven underwriting cycles and flight-to-quality dynamics consistent with the insurance-market evidence discussed in the literature.

Moreover, the model shows that ambiguity concerns affect both insurers' market valuation and their capital-management boundaries. Financial investment expands the range of aggregate capacity and changes the dynamics of underwriting cycles. In addition, when underwriting surplus and financial returns are sufficiently negatively correlated, the investment hedge can reduce equilibrium loadings and may generate negative insurance prices in high-capacity states. This suggests that observed underwriting losses need not simply reflect pricing mistakes; they may also arise from an equilibrium investment-driven pricing mechanism. Together, these findings offer a unified framework linking insurers' robustness concerns, financial frictions, investment opportunities, and market dynamics, with implications for robust pricing and liquidity management in the insurance industry.

\bibliographystyle{apalike} 
\bibliography{aguiar} 

\end{document}